\documentclass[12pt,preprint]{aastex}

\shorttitle{Sub-mJy USS radio sources}

\shortauthors{Afonso et al.}

\begin{document}

\title{Ultra Steep Spectrum radio sources in the Lockman Hole: SERVS identifications and redshift distribution at the faintest radio fluxes}

\author{
J. Afonso\altaffilmark{1,2},
L. Bizzocchi\altaffilmark{1,2},
E.~Ibar\altaffilmark{3},
M. Grossi\altaffilmark{1,2},
C. Simpson\altaffilmark{4},
S. Chapman\altaffilmark{5},
M.\,J.~Jarvis\altaffilmark{6},
H. Rottgering\altaffilmark{7},
R.\,P.~Norris\altaffilmark{8},
J. Dunlop\altaffilmark{9},
R.\,J.~Ivison\altaffilmark{3,9},
H. Messias\altaffilmark{1,2},
J. Pforr\altaffilmark{14},
M. Vaccari\altaffilmark{15},
N. Seymour\altaffilmark{10},
P. Best\altaffilmark{9},
E. Gonz\'{a}lez-Solares\altaffilmark{12},
D. Farrah\altaffilmark{11},
C.\,A.\,C. Fernandes\altaffilmark{1,2},
J.-S. Huang\altaffilmark{19},
M. Lacy\altaffilmark{13},
C. Maraston\altaffilmark{14},
L. Marchetti\altaffilmark{15},
J.-C. Mauduit\altaffilmark{16},
S. Oliver\altaffilmark{11},
D. Rigopoulou\altaffilmark{20},
S.\,A. Stanford\altaffilmark{17,18},
J. Surace\altaffilmark{16},
G. Zeimann\altaffilmark{17}
}

\altaffiltext{1}{Observat\'{o}rio Astron\'{o}mico de Lisboa, Faculdade de Ci\^{e}ncias, Universidade de Lisboa, Tapada da Ajuda, 1349-018
Lisbon, Portugal; jafonso@oal.ul.pt}

\altaffiltext{2}{Centro de Astronomia e Astrof\'{\i}sica da Universidade de Lisboa, Lisbon, Portugal}

\altaffiltext{3}{UK Astronomy Technology Centre, Royal Observatory, Edinburgh EH9 3HJ, UK}

\altaffiltext{4}{Astrophysics Research Institute, Liverpool John Moores University, Twelve Quays House, Egerton Wharf, Birkenhead CH41 1LD}

\altaffiltext{5}{Institute of Astronomy, University of Cambridge, Madingley Road, Cambridge, CB3 0HA, UK}

\altaffiltext{6}{Centre for Astrophysics, Science \& Technology Research Institute, University of Hertfordshire, Hatfield, Herts, AL10 9AB, UK}

\altaffiltext{7}{Leiden Observatory, Leiden University, Oort Gebouw, PO Box 9513, 2300 RA Leiden, The Netherlands}

\altaffiltext{8}{CSIRO Astronomy \& Space Science, PO Box 76, Epping, NSW, 1710, Australia}

\altaffiltext{9}{SUPA, Institute for Astronomy, University of Edinburgh, Blackford Hill, Edinburgh EH9 3HJ, UK}

\altaffiltext{10}{Mullard Space Science Laboratory, UCL, Holmbury St Mary, Dorking, Surrey, RH5 6NT, UK}

\altaffiltext{11}{Astronomy Centre, Dept. of Physics \& Astronomy, University of Sussex, Brighton BN1 9QH, UK}

\altaffiltext{12}{Institute of Astronomy, University of Cambridge, Madingley Road, Cambridge, CB3 0HA, UK}

\altaffiltext{13}{National Radio Astronomy Observatory, 520 Edgemont Road, Charlottesville, VA 22903, USA}

\altaffiltext{14}{Institute of Cosmology and Gravitation, University of Portsmouth, Dennis Sciama Building, Burnaby Road, Portsmouth, PO1 3FX, UK}

\altaffiltext{15}{Department of Astronomy, University of Padova, vicolo Osservatorio 3, 35122, Padova, Italy}

\altaffiltext{16}{Infrared Processing and Analysis Center/Spitzer Science Center, California Institute of Technology, Mail Code 220-6, Pasadena, CA 91125, USA}

\altaffiltext{17}{Department of Physics, University of California, One Shields Avenue, Davis, CA95616, USA}

\altaffiltext{18}{IGPP, Lawrence Livermore National Laboratory, 7000 East Avenue, Livermore, CA94550, USA}

\altaffiltext{19}{Department of Astrophysics, Oxford University, Keble Road, Oxford OX1 3RH, UK}

\altaffiltext{20}{Harvard-Smithsonian Center for Astrophysics, 60 Garden Street, Cambridge, MA 02138, USA}

\begin{abstract}

Ultra Steep Spectrum (USS) radio sources have been successfully used to select powerful radio sources at high redshifts ($z\gtrsim2$). Typically restricted to large-sky surveys and relatively bright radio flux densities, it has gradually become possible to extend the USS search to sub-mJy levels, thanks to the recent appearance of sensitive low-frequency radio facilities. Here a first detailed analysis of the nature of the faintest USS sources is presented. By using Giant Metrewave Radio Telescope and Very Large Array radio observations of the Lockman Hole at 610\,MHz and 1.4\,GHz, a sample of 58 USS sources, with 610\,MHz integrated fluxes above 100\,$\mu$Jy, is assembled. Deep infrared data at 3.6 and 4.5\,$\mu$m from the {\em Spitzer Extragalactic Representative Volume Survey} (SERVS) is used to reliably identify counterparts for 48 (83\%) of these sources, showing an average total magnitude of $[3.6]_{AB}=19.8$\,mag. Spectroscopic redshifts for 14 USS sources, together with photometric redshift estimates, improved by the use of the deep SERVS data, for a further 19 objects, show redshifts ranging from $z=0.1$ to $z=2.8$, peaking at $z\sim 0.6$ and tailing off at high redshifts. The remaining 25 USS sources, with no redshift estimate, include the faintest [3.6] magnitudes, with 10 sources undetected at 3.6 and 4.5\,$\mu$m (typically $[3.6]\gtrsim 22-23$\,mag, from local measurements), which suggests the likely existence of higher redshifts among the sub-mJy USS population. The comparison with the Square Kilometre Array Design Studies Simulated Skies models indicate that Fanaroff-Riley type I radio sources and radio-quiet Active Galactic Nuclei may constitute the bulk of the faintest USS population, and raises the possibility that the high efficiency of the USS technique for the selection of high redshift sources remains even at the sub-mJy level.

\end{abstract}


\section{INTRODUCTION}

High-redshift radio galaxies (HzRGs) are amongst the most luminous galaxies and seem to be associated with the most massive systems (e.g. \citealt{vanBreugel99,Jarvis01b,Willott03,Rocca04,DeBreuck05}\linebreak \citealt{Seymour07}). A number of studies have shown that their host galaxies may contain large amounts of dust and violent star formation at the 1000\,M$_{\sun}$\,yr$^{-1}$ level and also large gas and dust reservoirs (e.g. \citealt{Dunlop94,Ivison95,Hughes97,Ivison98,Papadopoulos00, Archibald01,Klamer05,Reuland03,Reuland04,Reuland07}; but see \citealt{Rawlings04}). The presence of such huge amounts of star formation at high redshift, and the fact that we know that powerful radio galaxies in the lower redshift Universe are hosted by massive ellipticals \citep{Best98,McLure00,McLure04} suggests that these HzRGs are the precursors of massive elliptical galaxies at low redshift \citep[e.g.][]{Sadler03,Dunlop03}. As a result, HzRGs have been used as beacons to identify over-densities in the distant Universe, i.e. proto-cluster environments at $z\sim2-5$ \citep[e.g.][]{Stevens03,Venemans07}. Identifying and tracing the evolution of HzRGs thus offers a unique path to study galaxy and large scale structure formation and evolution from the earliest epochs.

One of the most successful tracers of HzRGs relies on the relation between the steepness of the radio spectra and redshift \citep[e.g.][]{Tielens79,Chambers96}. Although an ultra-steep (radio) spectrum (USS; $\alpha\lesssim -1$ with $S\propto \nu^{\alpha}$) does not guarantee a high redshift, and the USS selection may actually miss a large fraction of HzRGs \citep[e.g.][]{Waddington99,Jarvis01c,Jarvis09}, a higher fraction of high-redshift sources can be found among those with the steepest radio spectra, and most of the radio galaxies known at $z>3.5$ have been found using the USS criterion \citep{Blundell98,DeBreuck98,DeBreuck00,Jarvis01a,Jarvis01b,DeBreuck02a,Jarvis04,Cruz06,Miley08}. A satisfactory explanation for this trend is still missing. One possibility is that the steeper radio spectral index arises from a combination of an increased spectral curvature with redshift, and the redshifting of a concave radio spectrum \citep[e.g.,][]{Krolik91} to lower radio frequencies. Another explanation, increasingly being supported by the observations, suggests that steeper spectral indexes are linked to radio jets expanding in dense environments, a situation one would more frequently encounter in proto-cluster environments in the distant Universe \citep{Klamer06,Bryant09,Bornancini10}. A recent study of submillimetre galaxies \citep{Ibar10}, known to be high-redshift ($z\sim 2$) massive objects linked to the galaxy formation process in proto-clusters, seems to support this latter hypothesis: while the average slope of their radio continuum emission is indistinguishable from that of local star-forming galaxies ($\alpha^{\rm 1400\,MHz}_{\rm 610\,MHz}=-0.75$, with standard deviation of 0.29), those sources with an active galactic nuclei (AGN) dominated mid-IR spectrum typically show steeper radio spectral indices ($\alpha^{1400}_{610} \lesssim -1.0$).

Until recently, widefield (over several tens or hundreds of square degree) and, consequently, only moderately deep radio surveys have been used to select and study the USS radio population; for example, the sources of \citet{DeBreuck00,DeBreuck02a,DeBreuck02b,DeBreuck04} do not have 1400-MHz flux densities below $\sim$10--15\,mJy. This raises the question of whether a significant population of apparently fainter USS sources either at even higher redshifts and/or with lower AGN power is being missed by these surveys. At sub-mJy levels the issue could actually become more complex, as the population mix of radio sources changes dramatically from that observed above a few mJy, with an increasingly large contribution from the evolving star-forming galaxy population \citep[e.g][]{JarvisRawlings04,Afonso05,Simpson06,Smolcic08,Wilman08}. With the recent appearance of sensitive low-frequency radio facilities such as the Giant Metrewave Radio Telescope (GMRT), and with LOFAR being commissioned, it has become viable to extend the search for and subsequent study of USS sources to the $\mu$Jy regime \citep[e.g.][]{Bondi07,Owen09,Afonso09,Ibar09}. Detailed studies of these sources are fundamental, but still lacking, as the deepest possible supporting multi-wavelength data are an absolute necessity.

The infrared (IR) regime is, in this respect, of paramount importance: not only are USS sources potentially at very high redshift, but they are also potentially dusty, which means they will be increasingly difficult to detect at optical wavelengths. Wide-area deep IR surveys, which have became possible over the last few years with the {\em Spitzer Space Telescope}, are now allowing to significantly improve the study of USS sources. Of particular relevance is the recent transition of {\em Spitzer} to the post-cryogenic, or ``warm'' mission, phase. This change has allowed much larger surveys with the two shortest wavelength channels (3.6 and 4.5\,$\mu$m) of the Infrared Array Camera (IRAC), than was sensible during the cryogenic portion of the mission. The {\em Spitzer Extragalactic Representative Volume Survey} \citep[SERVS --][]{Mauduit11} is such a wide-area survey, covering 18 deg$^2$ to $\approx 1-2$\,$\mu$Jy in both the [3.6] and [4.5] bands, over five of the best-studied extragalactic fields: the European Large Area ISO Survey fields N1 and S1, the XMM-Newton Large Scale Structure survey, the {\it Chandra} Deep Field South and the Lockman Hole.

All these fields were, or are being covered at radio wavelengths to very deep levels \citep[see][for details]{Mauduit11}. We have selected one of them, the Lockman Hole, for a first detailed analysis of the faintest USS sources, given the availability of uniform coverage at two radio wavelengths, 1.4\,GHz and 610\,MHz, the early availability of the entire set of SERVS data in this field and the existence of a large body of supporting multiwavelength data. This paper is organised as follows. In Section~\ref{sect:selection} we describe the radio data and the USS sample selection. The multiwavelength data considered in this work is detailed in Section~\ref{sect:data}, while in Section~\ref{sect:id} the identification of USS sources at IR wavelengths is presented together with some indications of the nature of these sources from IR colour-colour diagnostic plots. The results of spectroscopic redshift determinations and photometric redshift estimates are considered in Section~\ref{sect:z}, along with the implications for USS source characterization. Section~\ref{sect:modelsvhz} compares the results from this work with SKADS Simulated Skies radio source population models and addresses the efficiency of the USS criteria at sub-mJy radio fluxes for the detection of very high-redshift sources. Finally, our main conclusions are summarized in Section~\ref{sect:concl}.

Throughout this paper we adopt $H_0 =71\,$ km\,s$^{-1}$\,Mpc$^{-1}$, $\Omega_M = 0.27$, and $\Omega_\Lambda= 0.73$. All quoted magnitudes are in the AB system.

\section{USS sample selection \label{sect:selection}}

The Lockman Hole radio coverage at 610\,MHz and 1.4\,GHz from \citet{Ibar09} was chosen for the selection of USS radio sources. Besides the depth and resolution of these surveys (reaching an r.m.s.\ of 6\,$\mu$Jy\,beam$^{-1}$ at 1.4\,GHz, with a restoring beam of $4.3\arcsec \times 4.2\arcsec$ and 15\,$\mu$Jy\,beam$^{-1}$ at 610\,MHz, with a $7.1\arcsec \times 6.5\arcsec$ restoring beam), they have been reduced in a similar way, using the same procedures, and with source catalogues which are easily comparable to each other. Problems associated with matching sources observed at different resolutions at different frequencies, for example, or using different source detection algorithms, are thus minimised.

\begin{figure}
\scalebox{0.8}{
\rotatebox{0}{
\includegraphics{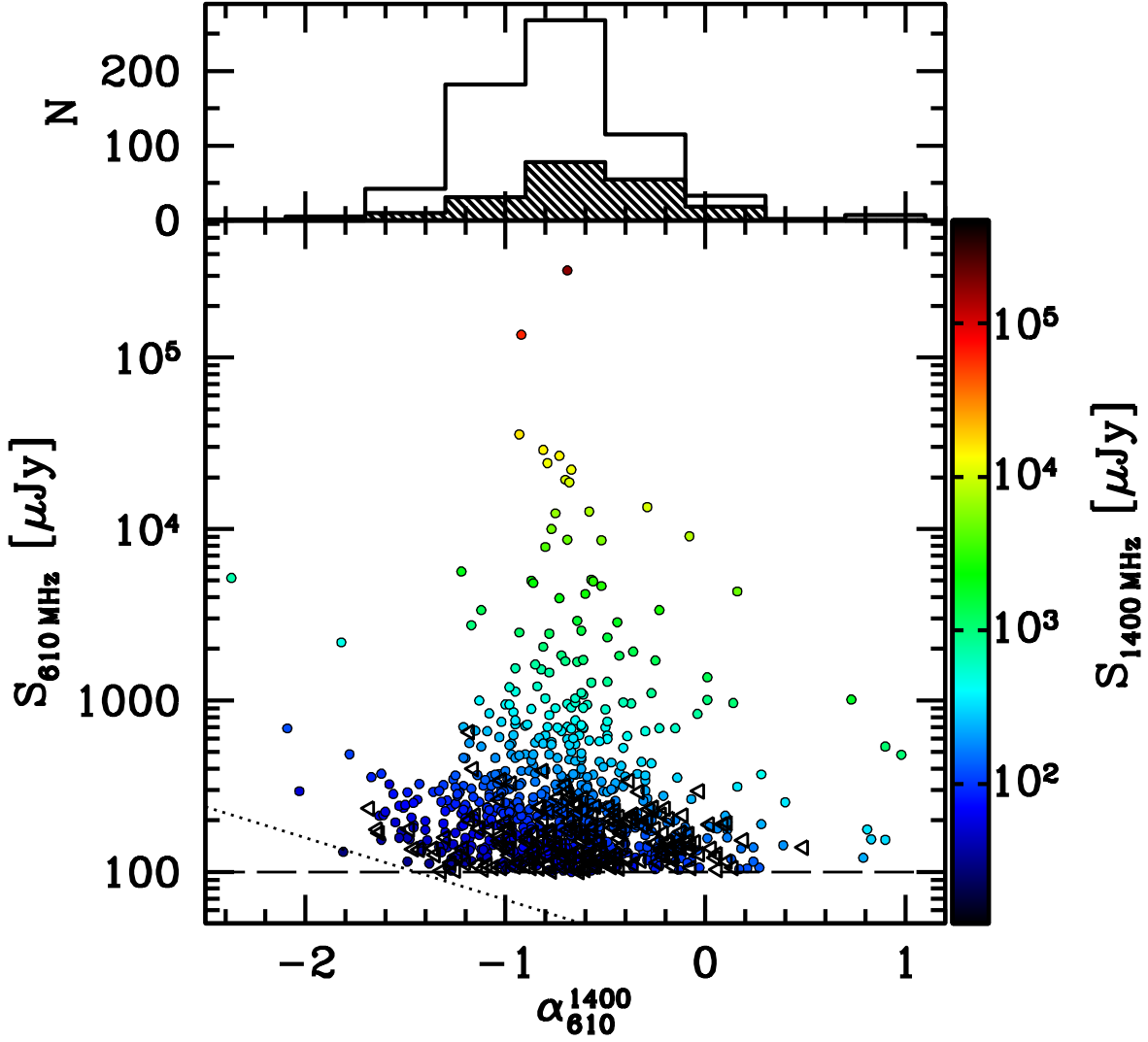}
}}
\caption{Flux density at 610 MHz {\it vs.} radio spectral index for faint radio sources in the Lockman Hole. Sources are colour coded according to their 1.4 GHz radio flux density (colour wedge on the right). Open symbols are upper limits in $\alpha$, representing sources with no 1.4 GHz detection. The dashed line represents the adopted lower limit for $S_{\rm 610\,MHz}$, while the dotted line shows the locus of points with $S_{\rm 1400\,MHz}=30\,\mu$Jy, corresponding to the \citet{Ibar09} $5\sigma$ peak flux limit at this frequency. Also shown (top) is the distribution of radio spectral indexes, with the hashed histogram representing $\alpha$ upper limits. The median value for the spectral index distribution (detections only) is -0.69.
\label{fig:alfaflux}}
\end{figure}

The area covered by both surveys is slightly different: the 1.4-GHz coverage with the VLA amounts to $\sim$0.56\,deg$^2$ while at 610\,MHz the GMRT coverage is nearly 1\,deg$^2$. In order to select USS sources we require coverage at both frequencies, and thus restricted our search to the smaller 1.4-GHz area. As indicated in \citet{Ibar09}, the 610-MHz catalogue contains 1,236 components within the area covered by the 1.4-GHz observations. After component merging \citep{Ibar09} we are left with 1,213 individual radio sources. In order to perform a more robust selection of USS sources we further restrict the sample in two ways: firstly, to avoid the unreliable spectral indices measurements of the faintest detected sources, we limit the 610\,MHz radio sample to integrated flux densities of $S_{\rm 610\,MHz}\geq 100$\,$\mu$Jy; secondly, we reject all sources at the edge (within $\sim$ 20 \arcsec) of the 1.4-GHz map, where the rapidly varying noise can bias the measured 1.4-GHz flux. This selection leaves a total of 928 (610-MHz selected) radio sources. After cross-correlation with the 1.4-GHz catalogue, we find spectral indices, $\alpha^{1400}_{610}$ (using $S_\nu \propto \nu^{\alpha}$, where $S_\nu$ represents the total integrated flux density at frequency $\nu$) for 662 of these, with an upper limit (non-detection at 1.4 GHz) obtained for 193 sources. We discard the remaining 73 sources, as they have an unreliable radio spectral index determination \citep[flagged entries in][]{Ibar09}. Figure~\ref{fig:alfaflux} represents the flux density at 610 MHz as a function of spectral index and the spectral index distribution for the sources in this field.


For the definition of the USS source sample we have considered commonly used parameters. These are by no means uniform, not in the frequencies used nor in the spectral index threshold -- e.g. $\alpha^{4850}_{151}< -0.981$ \citep{Blundell98}, $\alpha^{325}_{74}< -1.2$ \citep{Cohen04}, $\alpha^{1400}_{843}< -1.3$ \citep{DeBreuck04}, $\alpha^{1400}_{151}< -1.0$ \citep{Cruz06}, $\alpha^{843}_{408}< -1.0$ \citep{Broderick07}. In this work we have adopted a selection based on a conservative value of $\alpha^{1400}_{610}\leq -1.3$, which results in a sample of 65 USS sources, 10 of which are not detected at 1.4 GHz (i.e.\ only have an upper limit in $\alpha^{1400}_{610}$ based on a 5$\sigma$ threshold at 1.4\,GHz). Given the sensitivity of the USS criteria to the measured radio fluxes, we have visually inspected the radio images to identify and reject cases where the USS classification could be questionable. We thus rejected all cases where the comparison of the higher-resolution 1.4-GHz map with the 610-MHz map indicates that the radio emission at these frequencies traces different structures -- for example, when the 1.4-GHz map shows a pair of close neighbours that appear blended in the 610-MHz map. We have found 7 such cases. The final USS sample in the Lockman Hole thus consists of 58 sources (including 10 undetected at 1.4\,GHz), details of which are presented in Table~\ref{tab:sample}.

Given the flux measurement errors, the relative proximity of the radio frequencies employed, and the consequent relatively high spectral index uncertainties, it is important to characterize, in a statistical sense, the reliability of the USS sample. In particular, it is important to evaluate what is the expected ratio of non-USS sources that scatter into the USS regime due to measurement errors. We begin by considering an intrinsic spectral index distribution for 610\,MHz-selected radio sources described by a normal distribution of $\langle \alpha^{1400}_{610} \rangle \pm \sigma_\alpha = -0.80 \pm 0.33$ and a distribution for the spectral index measurement errors ($\Delta \alpha$) described by a normal distribution of $\langle \Delta \alpha \rangle \pm \sigma_{\Delta \alpha} = 0.30 \pm 0.12$. The latter is drawn directly from the observed measurement errors for 610\,MHz sub-mJy sources in the Lockman Hole and the former is chosen as to be compatible with the {\it observed} spectral index distribution for the sub-mJy radio sample \citep[see Table 7 in][]{Ibar09}. 
Under these assumptions, we find that a significant number of sources with {\em measured} spectral indices $\alpha^{1400}_{610}\leq -1.3$ are likely to have {\em intrinsic} spectral indices which are higher. This is not a surprise, but the expected behaviour from the selection of sources in the tail of the spectral index distribution \citep[see, for example,][]{DeBreuck04}, and can only be minimised by establishing a conservative cut in $\alpha$ and/or by trying to minimise the error affecting the $\alpha$ measurement -- by working at strong flux densities and/or by choosing widely separated radio frequencies for the measurement of $\alpha$. For the current work, the availability of quality radio data in the Lockman Hole and the objective of studying the faintest USS sources led to the choice of a conservative radio spectral index threshold ($\alpha^{1400}_{610}\leq -1.3$) for their selection. From the simulations described above, we expect only about a third of the sample in Table~\ref{tab:sample} to have an intrinsic radio spectral index below or equal to $-1.3$. However, most of the other sources can still be considered USS sources: around 75\% of the sample is expected to have an intrinsic radio spectral index of $\alpha^{1400}_{610}< -1.0$, still a commonly used USS selection threshold\footnote{We note that even increasing the spectral index measurement errors to $\langle \Delta \alpha \rangle \pm \sigma_{\Delta \alpha} = 0.40 \pm 0.12$ would still result in about two thirds of the selected sources having intrinsic radio spectral index of $\alpha^{1400}_{610}< -1.0$.}. This justifies our earlier conservative choice for the $\alpha^{1400}_{610}$ threshold, and indicates that the USS sample in Table~\ref{tab:sample} is appropriate for the statistical study of the faint radio USS source population -- keeping in mind that around 14 sources could have $\alpha^{1400}_{610}> -1.0$. In order to help identifying any bias from the expected fraction of non-USS sources (those with an intrinsic radio spectral index $\alpha^{1400}_{610}>-1.0$) included in the sample, we define a control sub-sample -- labeled as "A" in Table~\ref{tab:sample} and henceforward designated USS(A) subsample -- including only the 30 sources with the most secure USS classification.

We note that the assembled USS sample is not intended to be complete: the uncertainties in the spectral index determination will also scatter real USS sources into a non-USS classification. Completeness is nevertheless a requirement beyond the scope of this work.

\section{Optical and Infrared multi-wavelength data \label{sect:data}}

The Lockman Hole is one of the most extensively observed regions in the sky. Several surveys have obtained deep data on this field, from the X-rays to the radio. In this work, we make use of various multi-wavelength datasets for the identification of USS faint radio sources and the estimate of their redshift distribution. For the identification we use the IR data from SERVS \citep{Mauduit11}, which provides deep and uniform coverage of the entire region at 3.6 and 4.5\,$\mu$m. For the redshifts of these sources we use the spectroscopic data available on this field and complement it with photometric redshift estimates for sources with five or more broad-band photometric detections in the optical to IR ($0.36-4.5$\,$\mu$m) wavelength range. For this we use the ground-based optical imaging obtained with the Wide Field Camera at the Isaac Newton Telescope ($U g' r' i' Z$ bands), to 5$\sigma$ point source depths of 24.1, 24.9, 24.3, 23.6, 22.1\,mag (AB) respectively \citep{Gonzalez11}, the ground-based near-IR data, at $J$ and $K$ bands, from the UKIDSS Deep Extragalactic Survey \citep[DXS --][]{Lawrence07} reaching a limiting magnitude of $K_{AB} \sim 21.3$\,mag and the IR data from SERVS. Finally, we also use the {\em Spitzer} Wide-area Infrared Extragalactic survey \citep[SWIRE --][]{Lonsdale03} measurements at wavelengths of 5.8 and 8.0\,$\mu$m, to explore the infrared colours of the USS sources.
This vast body of wide-area multi-wavelength data was merged into a SERVS-selected multi-wavelength catalog, or SERVS Data Fusion, using a 1.5\arcsec\ search radius \citep{Vaccari11}. Deeper optical and near-IR observations of the Lockman Hole do exist, for parts of the field, e.g.\ Subaru imaging in the $r$, $i$ and $z$ bands \citep{Szokoly10}, and LBT observations at $U$, $B$ and $V$ \citep{Rovilos09}. These will be used for the study of specific USS sources in future papers, but the statistical study of the USS sample, collected over a wide-field, requires a more homogeneous approach which is facilitated by the wide-area datasets adopted here.


\section{The IR counterparts of USS faint radio sources \label{sect:id}}
\subsection{Identification rate}

Counterparts for the sample of USS sources were searched for in the SERVS images using the likelihood ratio method of \citet{Sutherland92}. Due to the better resolution delivered at the higher frequency, the nominal radio position was taken as the 1.4-GHz position, whenever a detection at this frequency was available (with the 610-MHz position adopted otherwise). The radio positional uncertainty was estimated following \citet{Ivison07}. We considered further that  3$\sigma$ positional uncertainties of less than 1$\arcsec$ are unrealistic given the possibility of systematic offsets \citep[e.g.,][]{Morrison10}. For each USS source, the 3.6\,$\mu$m band identification with the highest reliability \citep[${\mathcal R}$, see][]{Sutherland92} above 75\%  was taken as the real counterpart. The 4.5\,$\mu$m band image was inspected for identifications not present in the shorter wavelength band, but none were found. All identifications were inspected visually to check for special situations in which the likelihood ratio method would not apply, as in the case of non-independent sources (either in the radio or in the IR).

Using the deep SERVS observations 48 out of the 58 (83\%) USS sources possess an IRAC identification with ${\mathcal R} > 75$\%. The median AB magnitude for the USS source sample is $[3.6]=19.7$\,mag (an average value of $[3.6]=19.8$\,mag is found among detected counterparts). For the ten USS sources with no IRAC identification an estimate of the 5\,$\sigma$ [3.6]-magnitude lower limit has been estimated by using several measurements around the radio position. These results do not seem to be influenced by the contamination of the sample by non-USS sources. If we consider the more robust USS(A) sub-sample, we find an equally high IRAC identification rate of 80\%, and a very similar median magnitude for the counterparts of $[3.6]=19.6$\,mag (also an average of $[3.6]=19.6$\,mag considering only the detected counterparts). 

The identification rate for the USS sample using the infrared data in SERVS is comparable to that for the overall radio population in the Lockman Hole: 773 out of the 848 (91 per cent) sources with a radio spectral index determination or upper limit have an IRAC identification. It is important to note that chance coincidences have a limited impact in these high identification rates, as we have rejected all counterpart candidates with reliabilities below 75 per cent (and the vast majority of the identifications show very high reliabilities, ${\mathcal R}>95$ per cent). The false identification rate should thus be at the few per cent level. We have confirmed this result by running identification searches for random radio positions, obtained from the real USS sources positions (and assuming the same error regions) by adding a random distance between 30 and 45\arcsec\ along a random direction. The simulations returned random identification rates of between 2 and 4 per cent, as expected.

\begin{figure}
\scalebox{0.8}{
\rotatebox{0}{
\includegraphics{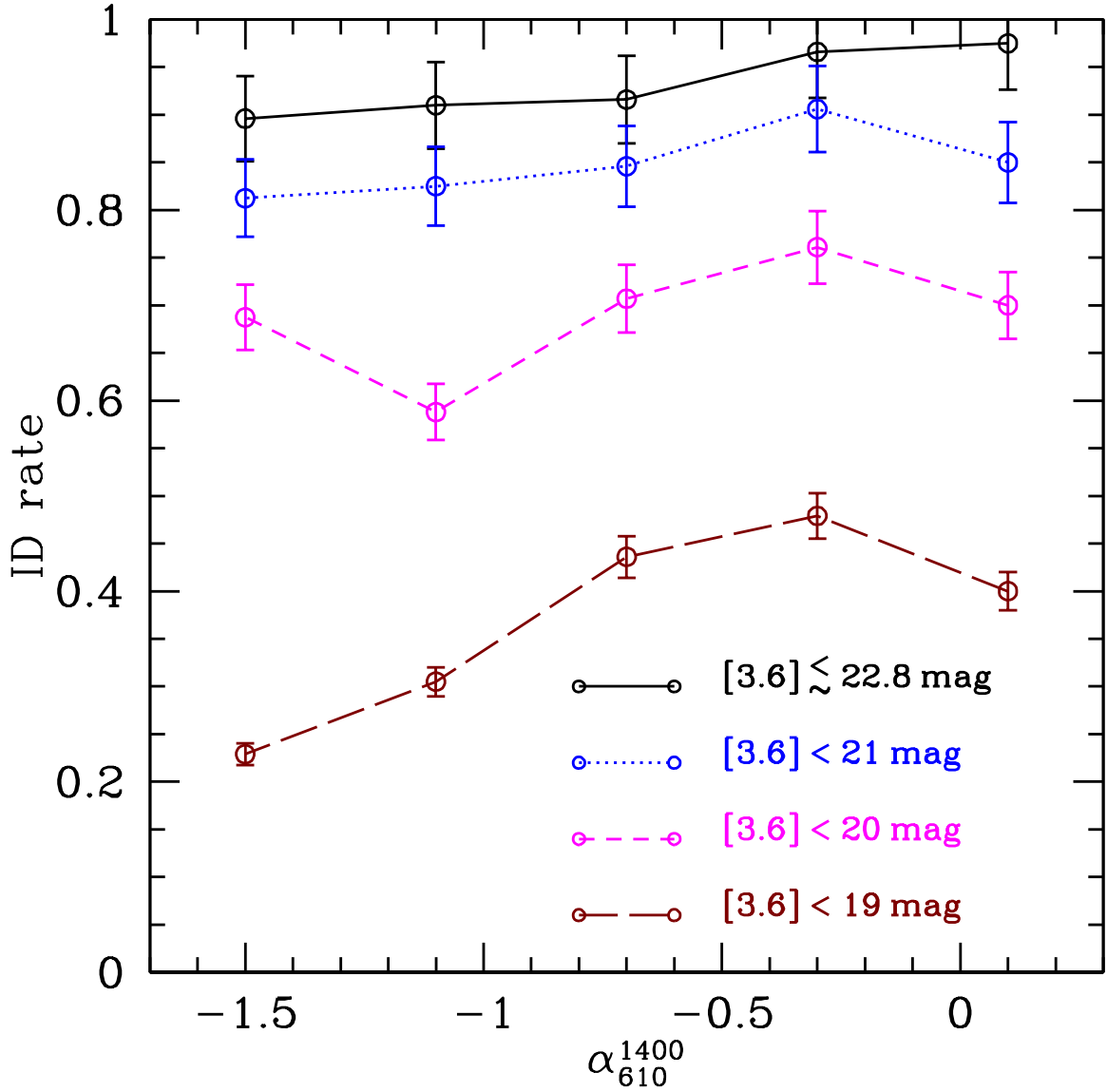}
}}
\caption{SERVS 3.6\,$\mu$m identification rate for faint radio sources in the Lockman Hole as a function of radio spectral index, for different [3.6]-band AB magnitude limits. The error bars are indicative of a 5\% error in the identification rate in each bin.
\label{fig:IDalfa}}
\end{figure}

The identification rate is also surprising in another way: when different radio spectral index ranges are considered, the identification rate always remains at the $\sim$90-per-cent level (Table~\ref{tab:ids}). This result is in contrast to previous work with USS sources at higher radio flux densities and brighter optical/IR magnitude limits, showing an identification rate which decreases with steeper radio spectral index \citep{Wieringa91,Intema10}. Such indications are usually taken to imply that sources with steeper radio spectral indices are, on average, at higher redshifts. Our high identification rate for the sub-mJy radio USS population does not, however, imply otherwise. The situation is akin to that of \citet{DeBreuck02a} which, by using relatively deep $K$-band observations (to $K\sim22$\,mag), achieve a near-IR identification rate of 94 per cent for their radio bright ($>10$\,mJy at 1.4\,GHz) USS sources. The dependence of the identification rate {\it vs.} radio spectral index relation on the optical/IR magnitude limit is clear from Figure~\ref{fig:IDalfa}. When the limiting magnitude for the infrared counterparts is set at $[3.6]=21$, 20 and 19\,mag not only does the identification rate show a significant decrease for all spectral indices but, by $[3.6]=19$\,mag, a more pronounced decrease of the identification rate of steeper spectrum sources seems to be settling in (23 per cent for $\alpha^{1400}_{610}\leq -1.3$ but $\sim$ 40 per cent for $\alpha^{1400}_{610}>-0.9$). This is a direct result of fainter [3.6] magnitudes for the sub-mJy USS population when compared to the flatter ($\alpha^{1400}_{610}>-0.9$) radio spectral index population: as mentioned above, the median magnitude for USS (measured $\alpha^{1400}_{610}\leq -1.3$) sources is $[3.6]=19.7$\,mag (average of $[3.6]=19.8$\,mag for detected counterparts) while amongst sources with a measured radio spectral index of $\alpha^{1400}_{610}>-0.9$ the median magnitude is $[3.6]=19.3$\,mag (average of $[3.6]=19.2$\,mag for detected counterparts). In parallel to the ``traditional'' (above the mJy level) USS population, this is also compatible with higher average redshifts among the steepest spectrum sources.

\begin{figure}
\scalebox{0.8}{
\rotatebox{0}{
\includegraphics{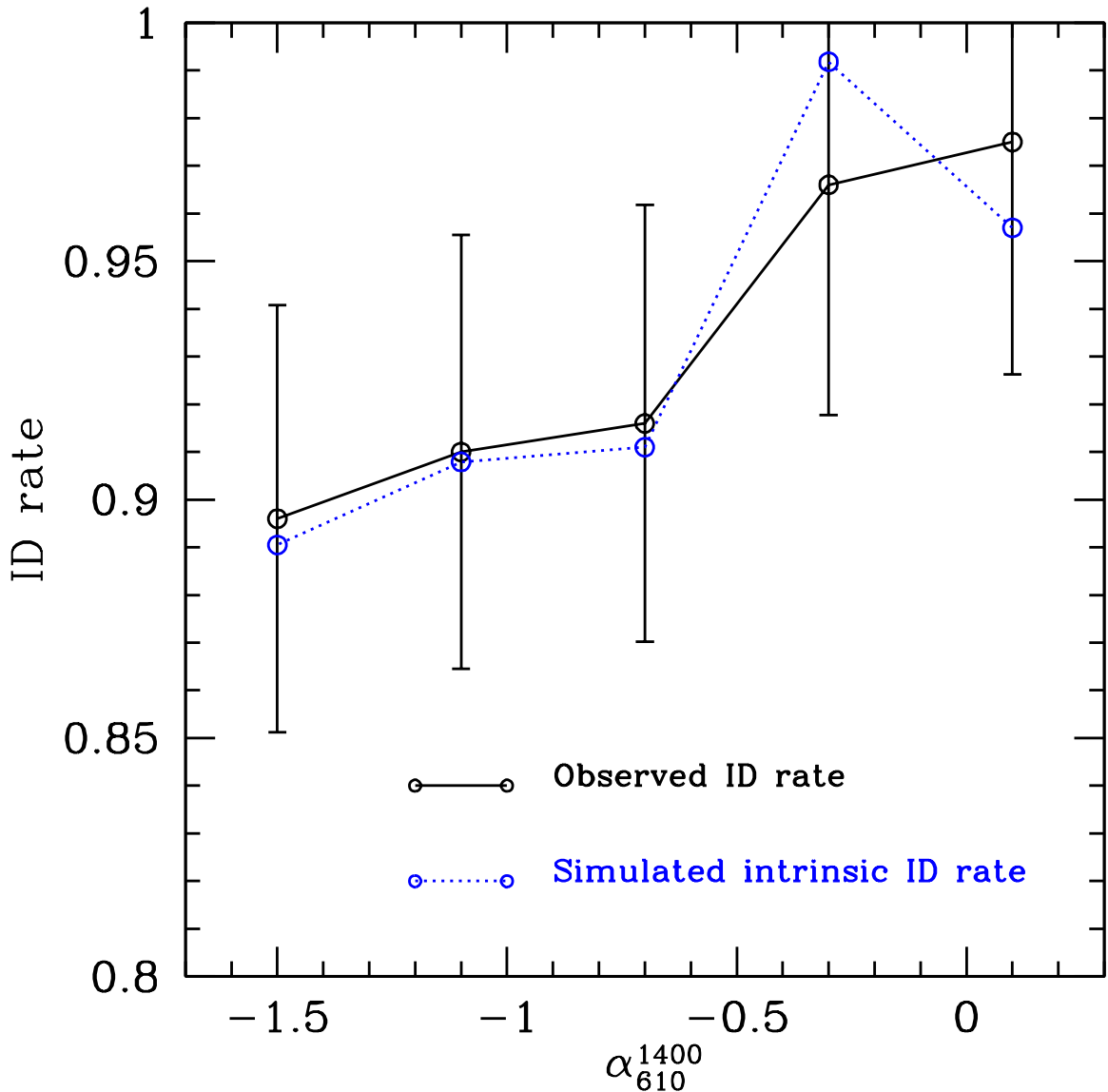}
}}
\caption{The observed SERVS 3.6\,$\mu$m identification rate for faint radio sources in the Lockman Hole as a function of radio spectral index (solid line) and a simulated intrinsic identification rate {\it vs} radio spectral index relation (dotted line) that would originate it after taking into account the measurement errors for the radio spectral index. This shows that the scatering of non-USS sources into the USS regime, due to the measurement errors, cannot explain the high identification rate observed for all spectral indices.
\label{fig:IDalfath}}
\end{figure}

The issue of the contamination of the USS sample by non-USS sources should also be considered here, as the spectral index measurement errors could affect the {\it observed} behaviour of the identification rate with spectral index. In general, measurement errors in the spectral index determination would be expected to flatten out any intrinsic correlation between identification rate and spectral index, if one exists, given the more significant scattering of presumably higher-identification rate non-USS sources (which exist in higher relative numbers) into the low-identification rate (and less populated) USS regime. The question then becomes how much of the almost flat behaviour of the identification rate with spectral index present in Figure~\ref{fig:IDalfa} is due to contamination of the USS sample by non USS sources. This can be tackled by simulating the effects of applying the distribution of spectral index errors in the current radio sample (considering only the sub-mJy regime, being the relevant one) to some intrinsic identification rate {\it vs.} radio spectral index relation. In Figure~\ref{fig:IDalfath} we show the intrinsic identification rate {\it vs.} radio spectral index relation which, for an intrinsic spectral index distribution of 610\,MHz-selected radio sources described by a normal distribution of $\langle \alpha^{1400}_{610} \rangle \pm \sigma_\alpha = -0.80 \pm 0.33$ and a distribution for the spectral index measurement errors ($\Delta \alpha$) given by a normal distribution of $\langle \Delta \alpha \rangle \pm \sigma_{\Delta \alpha} = 0.30 \pm 0.12$ (see section~\ref{sect:selection}), would result in the observed relation. Steeper identification rate {\it vs} radio spectral index relations have been explored but found incompatible with the observations. The effect of the spectral index measurement errors or, equivalently, the effect of the contamination of the USS regime by non-USS sources is, under these assumptions and for the purpose of the identification rate {\it vs} radio spectral index relation, virtually inexistent, and cannot be the reason for the flat curve in Figure~\ref{fig:IDalfa}. As detailed above, this should then be mostly a result of the depth of the infrared observations employed here.


\subsection{IR colour diagnostics for USS sources}

\begin{figure}
\plottwo{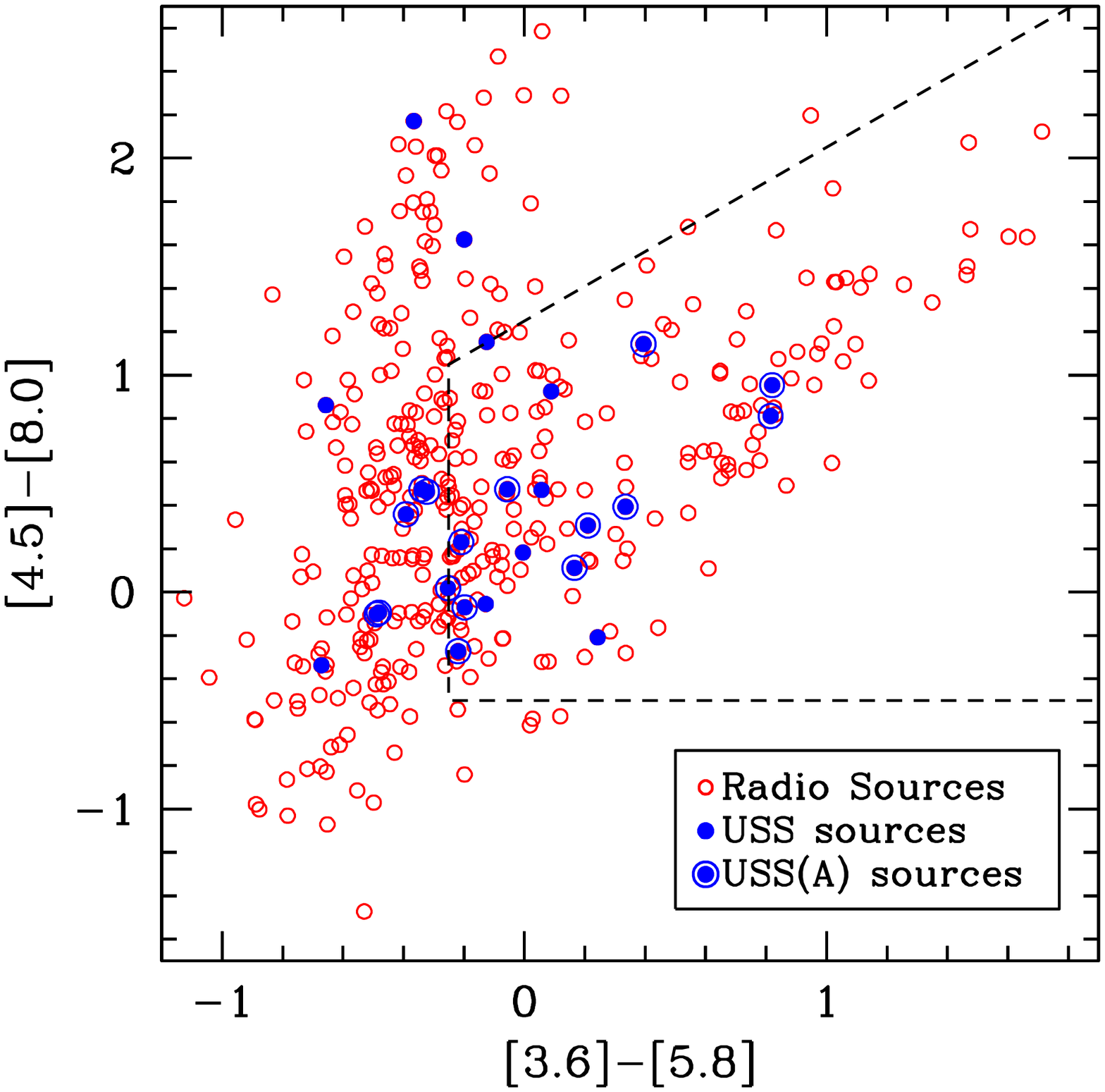}{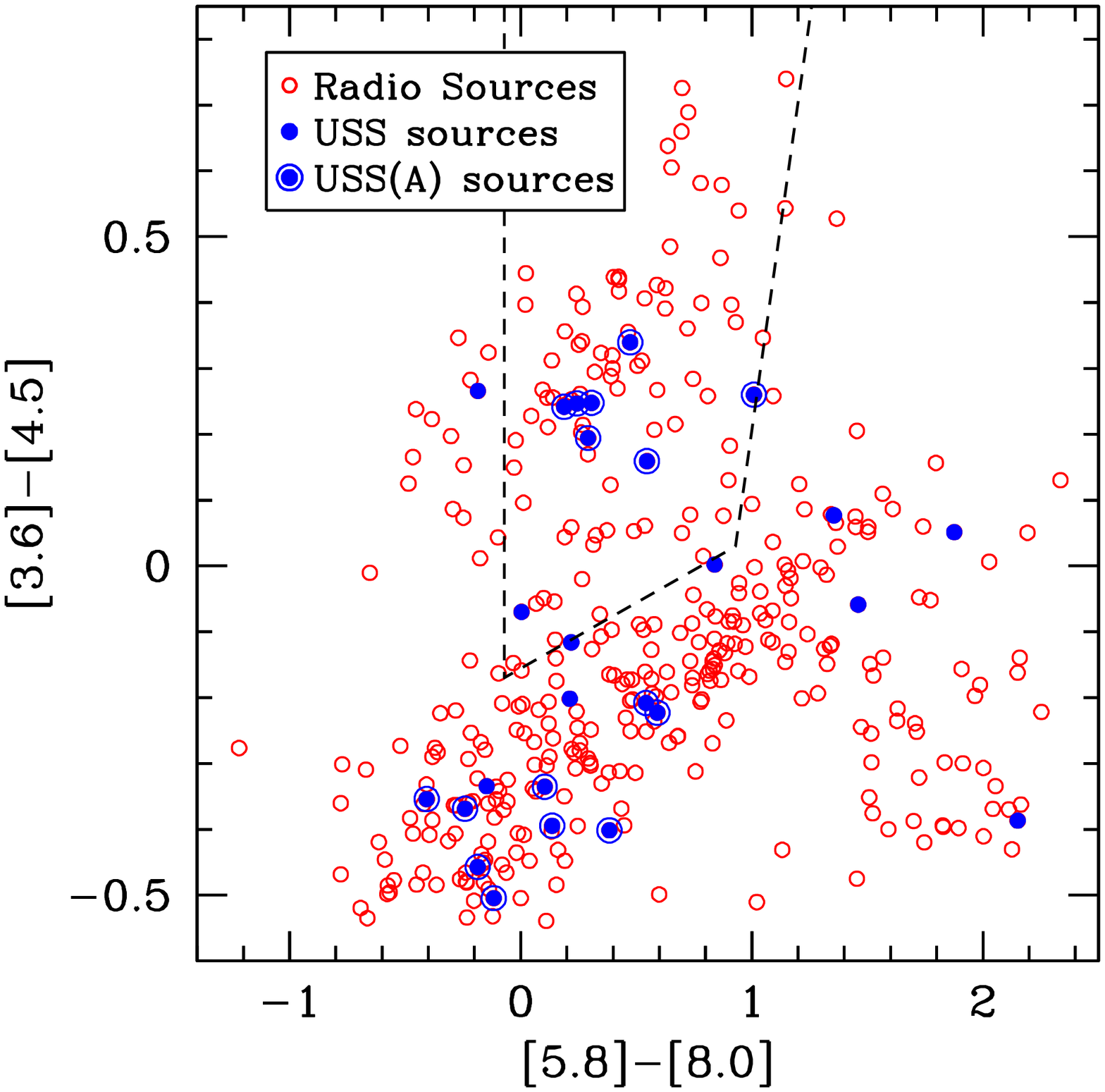}
\epsscale{0.5}
\plotone{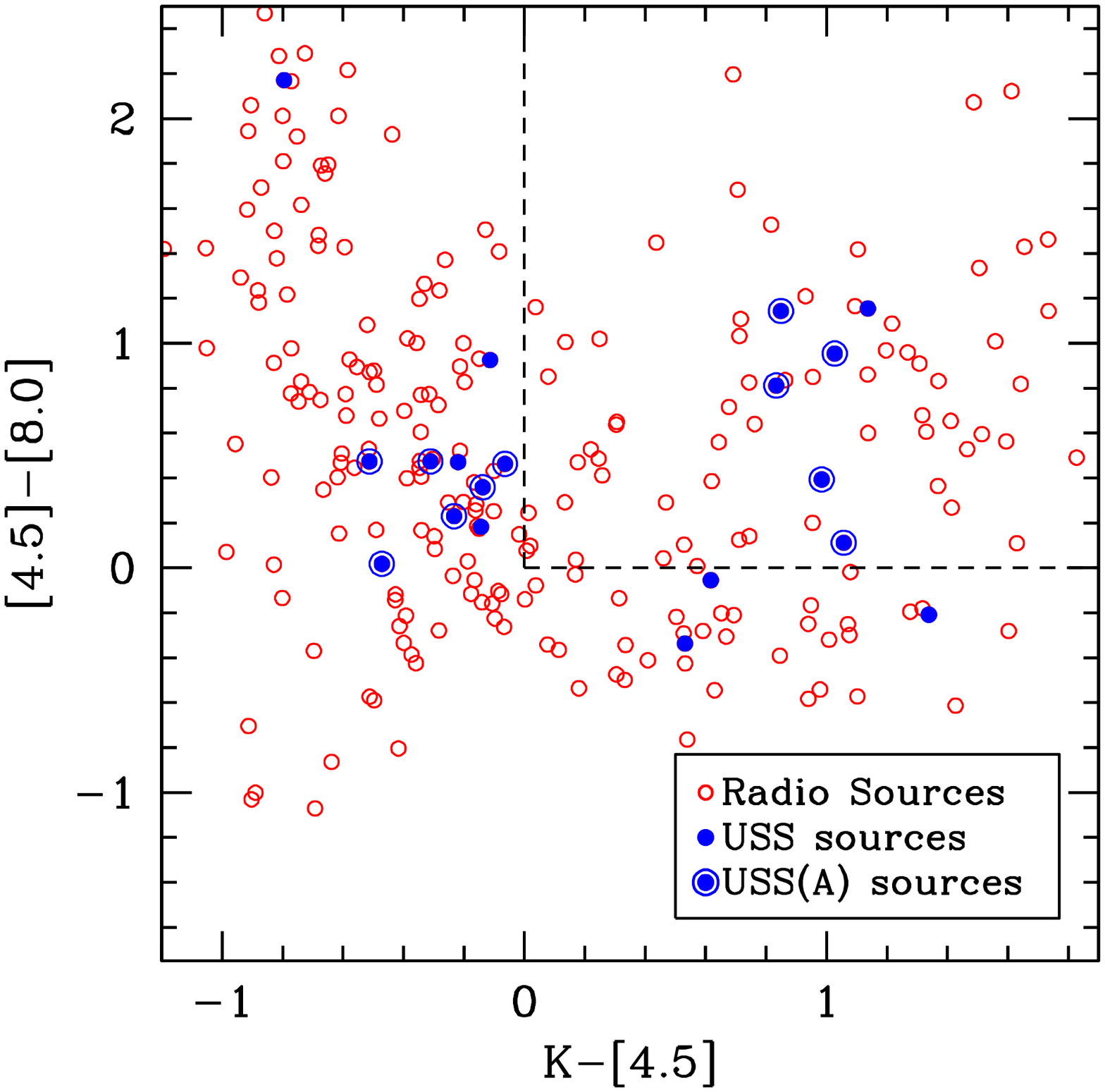}
\epsscale{1.0}
\caption{Infrared colour-colour diagnostics for radio sources in the Lockman Hole. Two of the most commonly used are shown, that of \citet{Lacy04} (top left) and that of \citet{Stern05} (top right), together with the recent KI diagram of \citet{Messias11}. The region delimited by the dashed line is the AGN region. Filled points denote the USS sample (outlined the USS(A) subsample), overlaid on the general sub-mJy radio population in the Lockman Hole.
\label{fig:lacystern}}
\end{figure}

Given the {\em Spitzer} coverage of the Lockman Hole region, both with SERVS and SWIRE, IR colours are available for many of the USS sources. This allows for the use of IR colour diagnostics plots, which may offer further clues about the nature of these sources. Figure~\ref{fig:lacystern} shows the location of radio sources in the Lockman Hole with appropriate IR detections in two of the most commonly used IR diagnostic diagrams \citep{Lacy04,Stern05} together with the recent KI diagnostic diagram of \citet{Messias11}. USS sources show a wide range of IR colours, with a significant fraction being associated with AGN colours -- among those USS sources with simultaneous detections at the relevant infrared bands about two-thirds fall in the AGN region for the Lacy et al.\ criteria, while around a third do so for the Stern et al.\ and the Messias et al.\ criteria. In spite of the small numbers of sources with appropriate photometry, the USS(A) subsample does not appear to differ significantly from the full USS sample. While these criteria are known to be contaminated by non-AGN galaxies, in particular at high ($z\gtrsim 2.5$) redshifts, and are certainly not complete \citep[i.e., AGN sources do appear outside the designated regions, see][]{Donley08,Barmby08}, they provide a strong indication that the USS population contains a substantial amount of IR-detected AGN. We note, nevertheless, that different accretion modes may be found in radio selected AGN, and in particular a radiatively inefficient ``radio mode'' may not be detectable at IR wavelengths \citep[e.g.,][]{Croton06,Hardcastle07,Tasse08,Griffith10}.

\section{Redshift distribution \label{sect:z}}
\subsection{Spectroscopic redshifts}

Several spectroscopic observations have targeted specific classes of sources in the Lockman Hole in the past \cite[e.g.][for X-ray and sub-millimetre sources, respectively]{Lehmann01,Chapman05}, but, besides the SDSS coverage of the brighter optical sources, only recently extensive deep spectroscopic surveys have been initiated. Among the 58 USS sources in this work, we have found two spectroscopic redshift determinations from published follow-up observations of X-ray sources in the ROSAT Deep Survey \citep{Schmidt98,Lehmann01}, six from an ongoing MMT-hectospec spectroscopic survey of SWIRE 24\,$\mu$m sources \citep{Huang10}, and four were found among a recent Keck spectroscopic survey of faint radio sources in this field (P.I.\ Scott Chapman). Finally, two sources had recent mid-infrared spectroscopy observations obtained with the {\em Spitzer}-InfraRed Spectrograph (IRS), as part of an analysis of candidate AGN-dominated sub-millimetre galaxies \citep{Coppin10}. These observations originate the two current highest spectroscopic redshifts in the USS sample, at $z=2.56$ and $z=2.76$.  Table~\ref{tab:spectz} shows details for these 14 sources.

\subsection{Photometric redshifts}

For the remainder of the USS sample it is possible to obtain photometric redshift estimates. Some of the sources have had a photometric redshift estimate from the SWIRE survey, from \citet{RowanRobinson08}. The procedure there employs both galaxy and quasi-stellar object templates applied to data at $0.36-4.5$\,$\mu$m and a set of four IR emission templates fitted to IR-excess data beyond 3.6\,$\mu$m. We find a total of 12 objects with a photometric redshift derived using 5 or more photometric bands, with 5 objects overlapping the current spectroscopic redshift sample. However, near-IR data, unavailable at the time for the Lockman Hole, could not be used by \citet{RowanRobinson08}. Given the recent availability of data from the UKIDSS Deep Extragalactic Survey on the Lockman Hole and the new, deeper, SERVS data, we have included these photometric bands and used both EAZY \citep{Brammer08} and HyperZ \citep[][following the procedure outlined in \citealp{Maraston06}]{Bolzonella00} to find photometric redshifts estimates for a larger number of sources. We use aperture corrected magnitudes measured in a radius of 1.6\arcsec, 2.0\arcsec\ and 1.9\arcsec\ for INT WFC, UKIDSS and IRAC respectively. The strength of EAZY's template error function was adjusted to provide the best fit to the existing spectroscopic data; although few objects have reliable spectroscopic redshifts, the chosen value was very similar to that used by \citet{Simpson10} for the 100-$\mu$Jy 1.4\,GHz radio sample of \citet{Simpson06}.

\begin{figure}
\scalebox{0.8}{
\rotatebox{0}{
\includegraphics{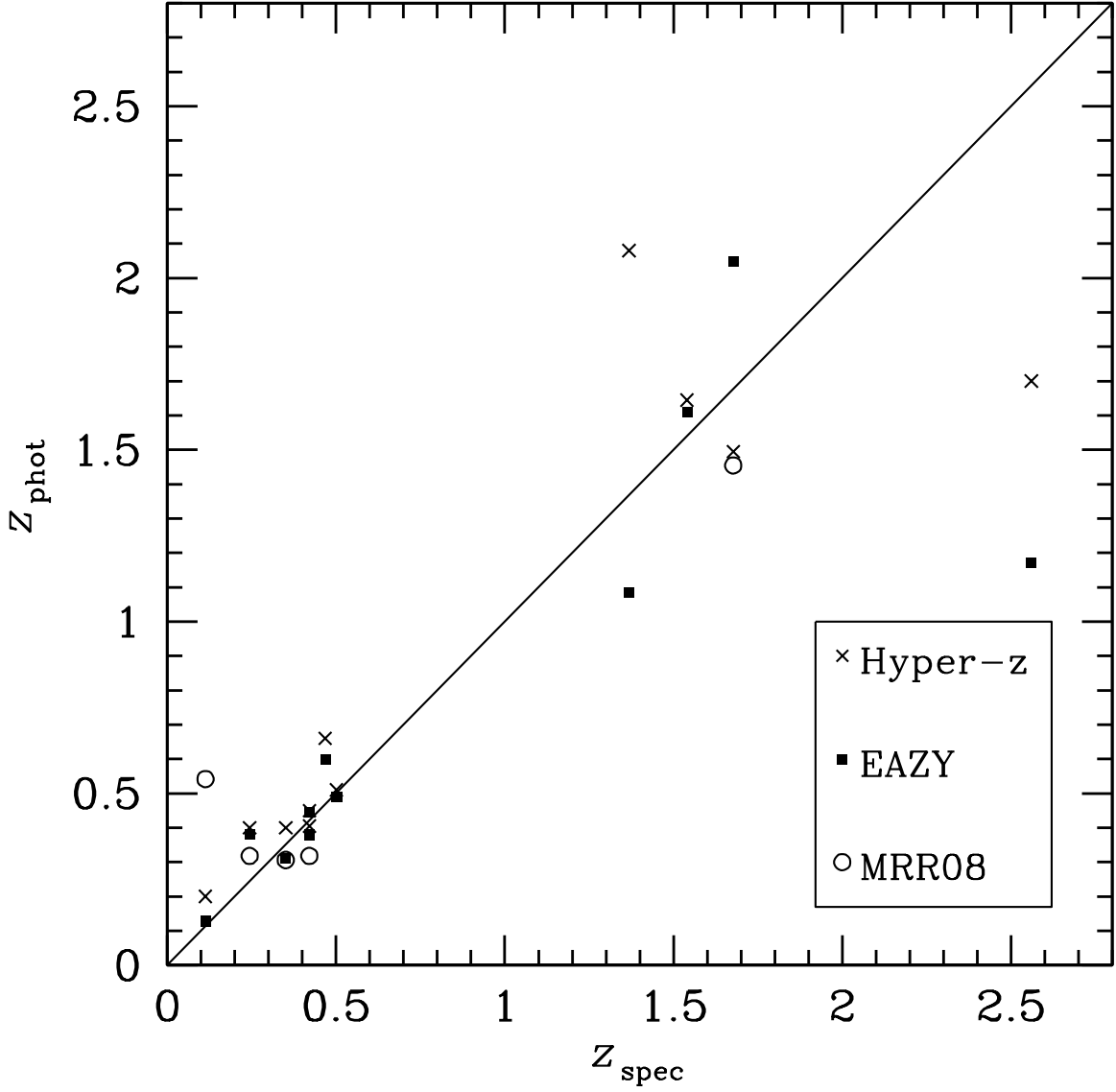}
}}
\caption{Comparison between spectroscopic redshifts and photometric redshifts obtained through three different procedures. See Section~\ref {sect:z} for details.
\label{fig:spectphotz}}
\end{figure}

Accepting only those photometric redshifts resulting from 5 or more photometric measurements in the $0.36-4.5\,\mu$m range, we obtain estimates for 30 sources, 11 of which have an available spectroscopic redshift measurement. The comparison between photometric and spectroscopic redshifts is indicated in Figure~\ref{fig:spectphotz}, and shows a good agreement out to $z\sim 2$.  The performances of EAZY and HyperZ are similar, with $\sigma _{nmad}$ \citep[using the normalized median absolute deviation as defined in][]{Brammer08} of $|\Delta z|/(1+z_{\mathrm{spec}})$ of 0.06 and 0.07, respectively. There is one ``catastrophic'' outlier ($|\Delta z|/(1+z_{\mathrm{spec}})>5\,\sigma_{nmad}$) for the photometric redshift of LH610MHzJ105407.0+573308 as given by EAZY ($|\Delta z|/(1+z_{\mathrm{spec}}) = 6.3\,\sigma_{nmad}$). This source has the second highest spectroscopic redshift of the USS sample, $z_{\mathrm{spec}}=2.56$, a value obtained from IRS mid-infrared spectroscopy. A detailed inspection of the resulting photometric redshifts probability distributions for this source shows broad and flat distributions between $z\sim 1$ and $z\sim 3$, with high probability secondary solutions at $z\sim 2.3-2.4$.  This highlights the unavoidable limitations of photometric redshift estimates for some sources, in particular at higher redshifts. In Table~\ref{tab:sample} and in what follows we adopt the HyperZ estimates, noting there would be no significant change by the adoption of the photometric redshifts from EAZY.

\begin{figure}
\scalebox{0.8}{
\rotatebox{0}{
\includegraphics{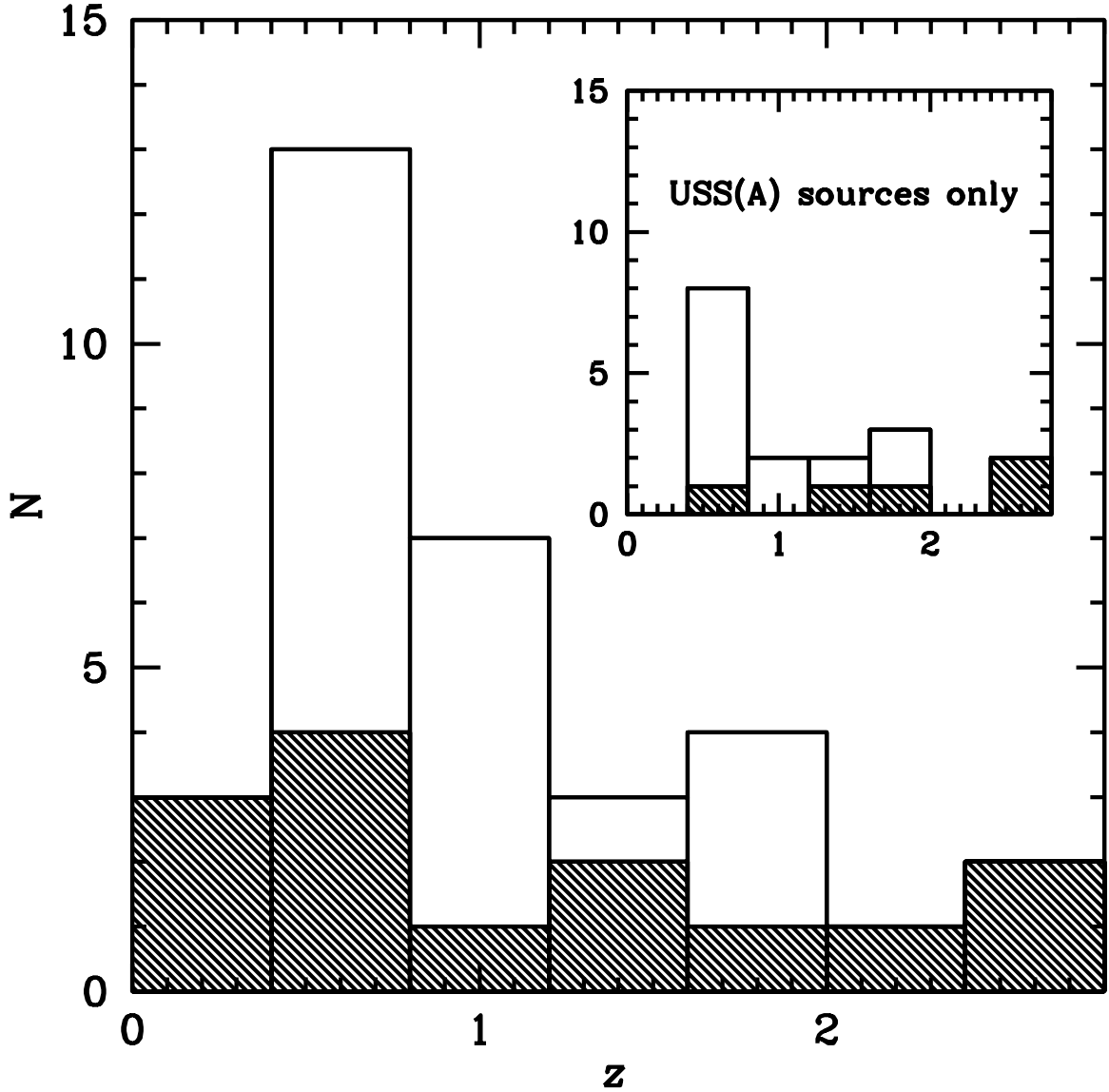}
}}
\caption{Redshift distribution for radio-faint USS sources in the Lockman Hole. Filled histogram denotes sources with a spectroscopic redshift determination, while the open region refers to photometric redshift estimates. A further 25 USS sources (43\% of the full sample) exist but with no redshift estimate, mostly at fainter [3.6] fluxes and likely to be found at higher redshifts. The inset shows the redshift distribution for the USS(A) subsample.
\label{fig:z-distr}}
\end{figure}

The redshift distribution for the sub-sample of USS sources with a redshift determination is presented in Figure~\ref{fig:z-distr}. It shows a significant presence of high redshift sources, with a peak at $z\sim 0.6$ and extending beyond $z\sim2$. The redshift distribution appears to be unchanged if one considers only the USS(A) subsample (inset in Figure~\ref{fig:z-distr}), which points to a limited effect of the small number of real non-USS sources expected to be included in this sample.

The radio monochromatic luminosities for these sources vary between \linebreak $L_{\rm 610\,MHz}\sim 10^{21.9}$\,W\,Hz$^{-1}$ and $L_{\rm 610\,MHz}\sim 10^{26}$\,W\,Hz$^{-1}$: while the lower values could be explained by lower power AGN or even star-forming galaxies \citep[e.g.][their Figure 7]{Afonso05} -- although the latter would pose problems for the interpretation of the steepness of the radio spectrum -- the higher values are clear indications of powerful AGN activity, already above the FRI/FRII luminosity break. Although beyond the scope of the present paper we should stress other clear indications for the existence of powerful AGN in the USS sample -- two of the spectroscopic redshifts were obtained by follow-up observations of X-ray sources in the ROSAT Deep Survey \citep{Schmidt98,Lehmann01}, and show broad optical lines which classify them as type I AGN; also, as mentioned above, two other sources have been observed by {\em Spitzer}-IRS as candidate AGN-dominated sub-millimetre galaxies, and SED fitting to their mid-infrared emission indeed indicates substantial AGN contributions over this wavelength range \citep{Coppin10}.

\begin{figure}
\scalebox{0.8}{
\rotatebox{0}{
\includegraphics{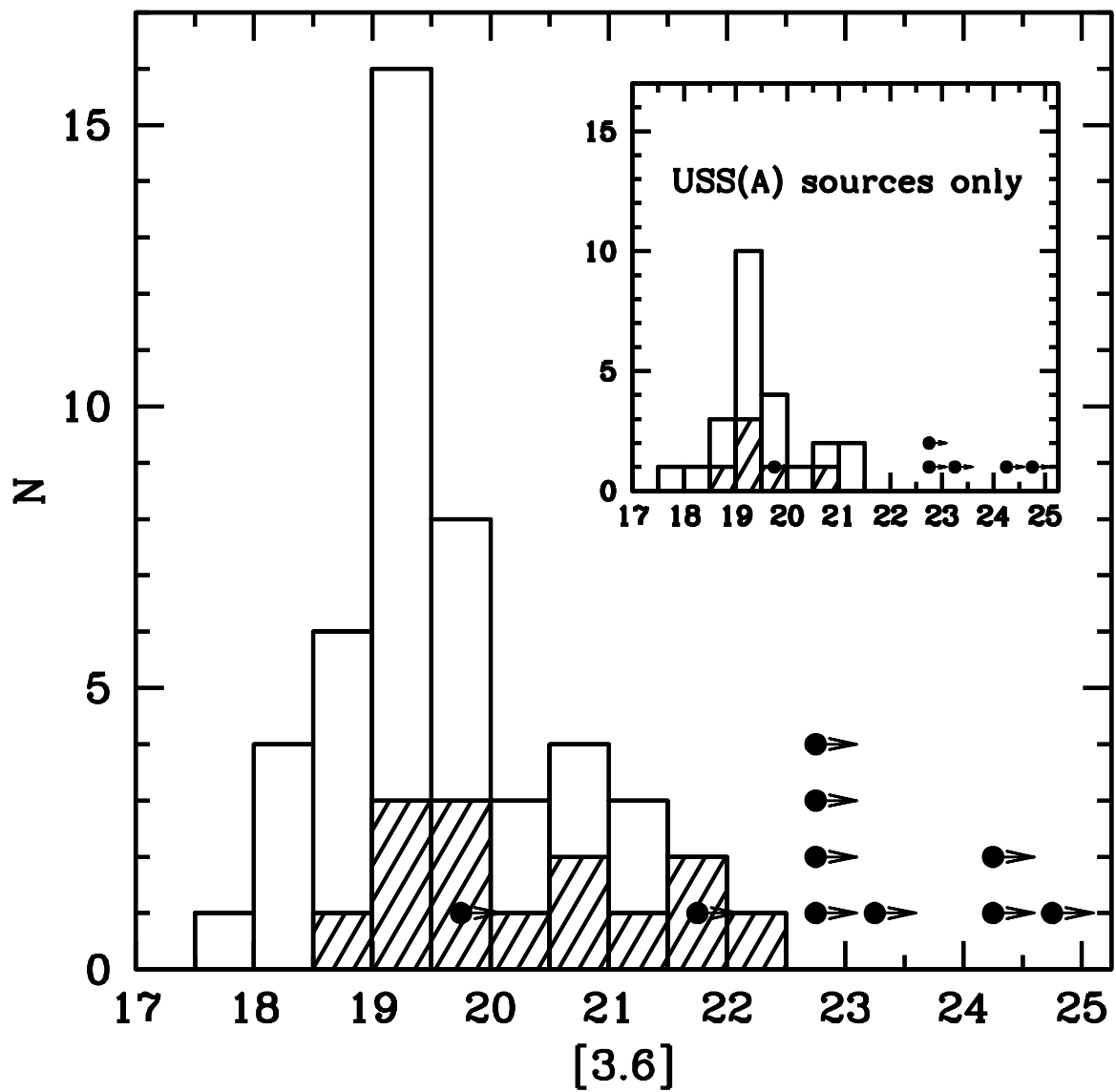}
}}
\caption{[3.6]-band AB magnitude distribution for USS sources in the Lockman Hole. Sources with no available redshift estimate are indicated by the filled histogram. The filled circles and attached arrows indicate the USS sources with no IRAC detection, placed at the bin location of their 5\,$\sigma$ [3.6] magnitude lower limit. The inset shows the magnitude distribution for the USS(A) subsample.
\label{fig:nonidmags}}
\end{figure}

One should recall that a redshift estimate exists for only 33 of the 58 USS sources, leaving more than 40 per cent of the sample with no redshift information. Figure~\ref{fig:nonidmags} shows that these objects correspond to the faintest [3.6] magnitudes, being potentially the highest redshift sources in our sample\footnote{One should also note from Figure~\ref{fig:nonidmags} that the [3.6]-band distribution of USS sources does not show any significant difference to that of the more restricted USS(A) subsample.}. In particular, the 10 USS sources with no IRAC detection are particularly interesting candidates for very high redshift objects. For these sources, the ratio of radio to 3.6\,$\mu$m fluxes can easily be above 50. Such USS sources are then also classified as Infrared-Faint Radio Sources (IFRS), a rare class of object first identified in the Australia Telescope Large Area Survey \citep{Norris06}. Recent follow-up studies have provided strong indications that the IFRS population probably contains a majority of obscured, high-redshift, radio-loud galaxies \citep{Norris07,Middelberg08,Huynh10,Norris10}. Interestingly, \citet{Middelberg10}, by studying the radio spectral indices of IFRS, also finds a significant overlap with the USS population (6 out of 17 IFRS were found to be USS sources). This argues for the likelihood of high redshifts for the 10 USS sources in the Lockman Hole without IRAC-SERVS counterparts.


\section{Comparison with SKADS Simulated Skies radio population models \label{sect:modelsvhz}}

While investigating the nature of USS radio sources at sub-mJy radio flux levels, and comparing it with the better known population revealed at brighter radio fluxes (above tens or hundreds of mJy), it is of the utmost interest to consider the predictions from radio population models. Unfortunately, radio spectral index information is usually not robustly included in such models, a result of a severe lack in understanding the physical origin of radio spectral indexes that differ from the nominal $\alpha\sim -0.7$ synchrotron value. Keeping such limitation in mind, we use the Square Kilometre Array Design Study (SKADS) Simulated Skies ($S^3$) simulations \citep{Wilman08} for two distinct situations: (a) Survey Model I - a radio survey reaching a detection sensitivity of 100\,$\mu$Jy at 610\,MHz over 0.6 square degree, which will be directly comparable with the current work; (b) Survey Model II - a radio survey reaching a detection sensitivity of 10\,mJy at 610\,MHz over 50 square degree - a situation more akin to the typical USS source search above the tens of mJy level. The simulations predict the number of sources, and respective redshifts, revealed by such surveys, among the following source types: radio-quiet (RQ) AGN, radio-loud AGN (including both FRI and FRII types), Gigahertz Peaked Sources (GPS) and  star-forming galaxies.

\begin{figure}
\plottwo{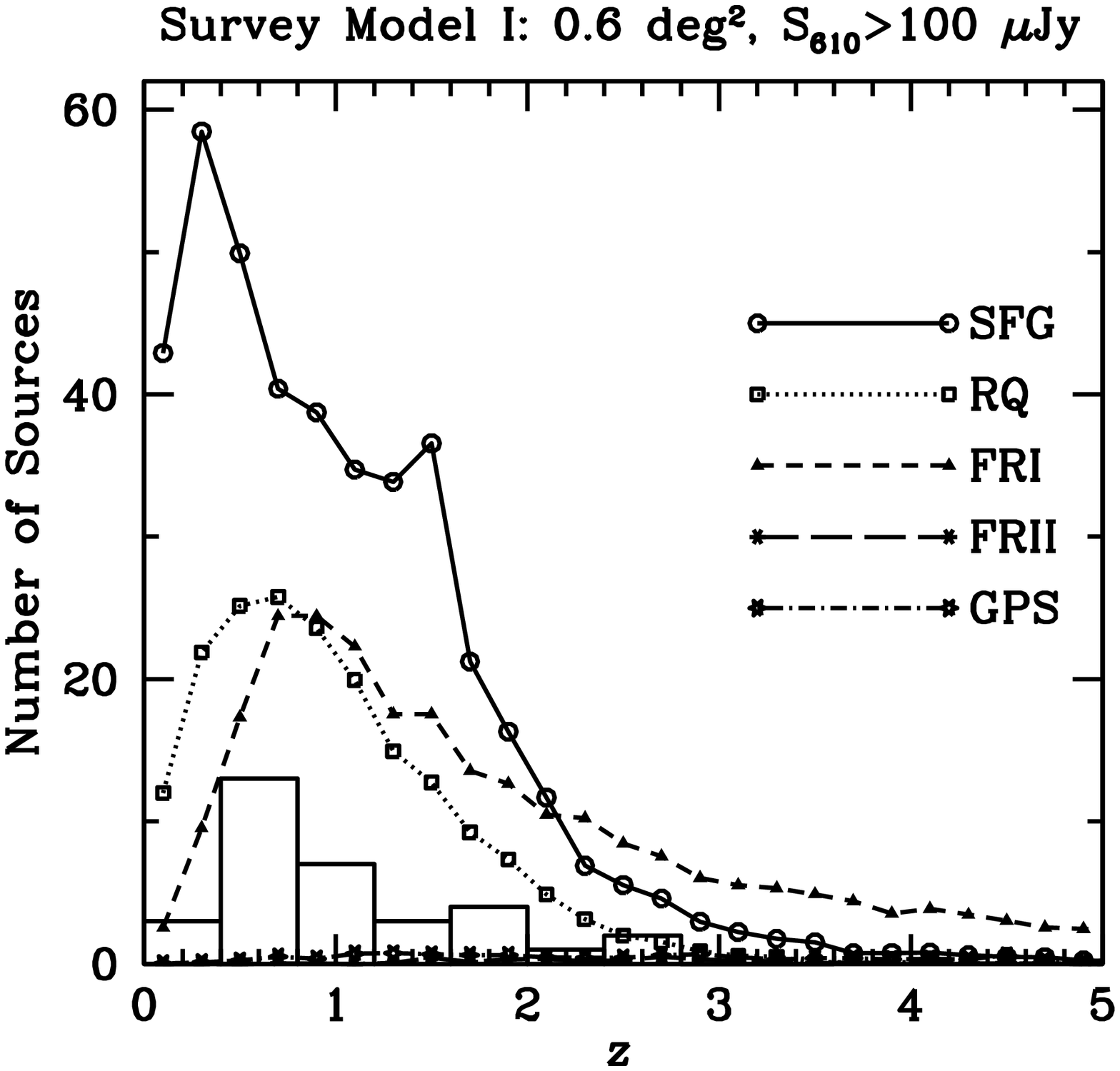}{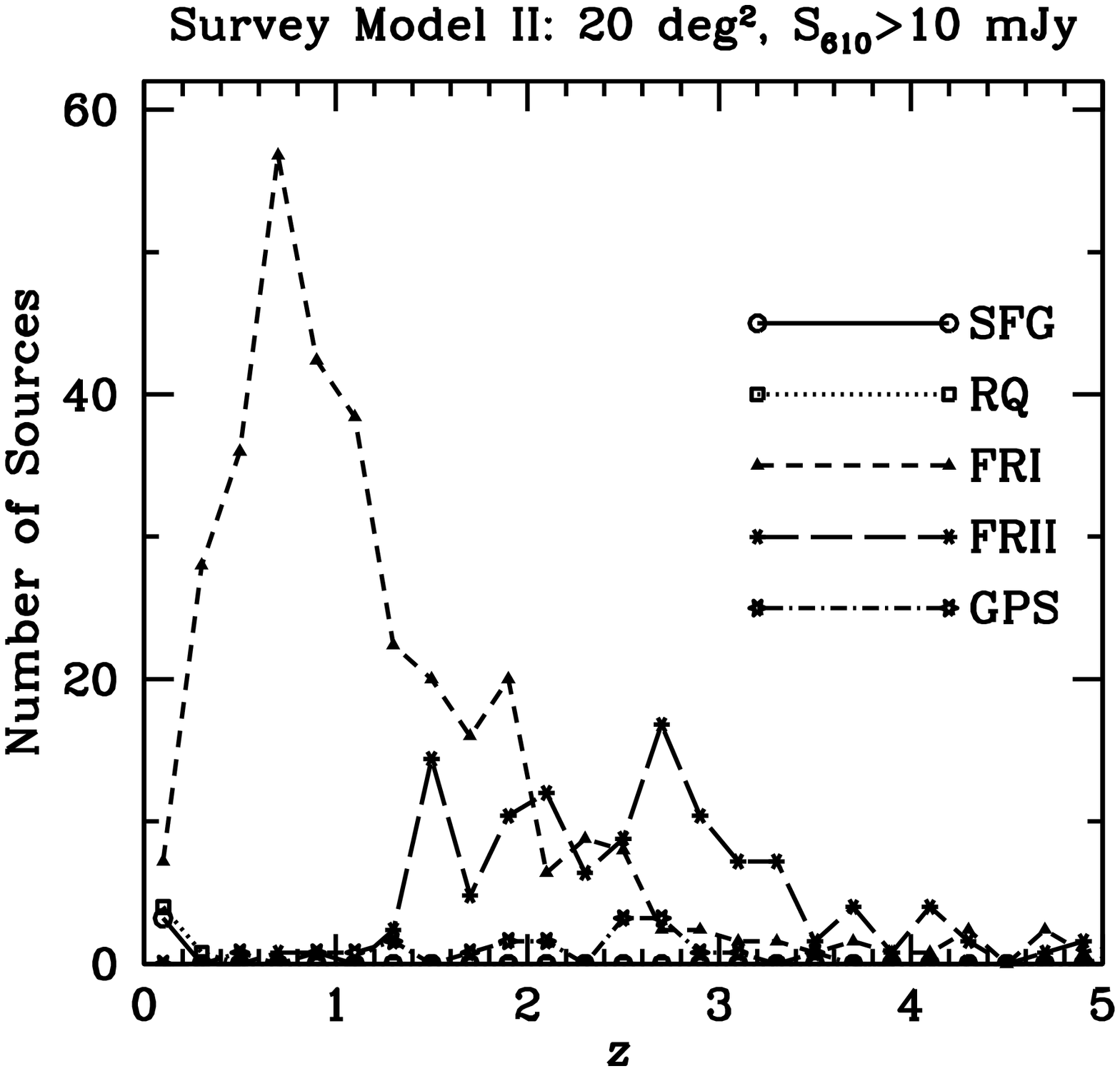}
\caption{Predictions from the SKADS Simulated Skies models for the redshift distributions of radio source populations, irrespective of their radio spectral indices, for two different types of radio survey: Survey Model I (left) is a radio survey reaching a detection sensitivity of 100\,$\mu$Jy at 610MHz over 0.6 square degree, similar to the Lockman Hole radio survey considered in the current work; Survey model II (right) is representative of a wider (50 square degree) and shallower (10\,mJy detection limit) survey, comparable to surveys frequently used to search for USS sources with radio flux densities above the tens of mJy level. The plot for Survey Model I, on the left, also displays the observed redshift distribution for USS sources in this work (open histogram), reproducing Figure~\ref{fig:z-distr} for a more straightforward comparison.
\label{fig:models}}
\end{figure}

In Figure~\ref{fig:models}, the redshift distributions for sources detected in both Survey Models are represented. The major obvious difference is the source population which the USS criteria will act upon. At high flux densities, the major contributors to the source counts will be powerful AGN of the FRI and FRII types, the latter dominating above $z\sim2$. When sampling sub-mJy radio flux densities, star-forming galaxies along with lower-luminosity AGN dominate the source detections, with FRI and RQ sources contributing significantly at $z<2$. At higher redshifts ($z>2$), FRIs dominate the survey model detections, however this population is very poorly constrained at such redshifts.

The $S^3$ models are compatible with the current work, and also with previous investigations at higher radio flux densities. If indeed the USS criteria is sensitive to radio jets in high density environments, it will pick up a fraction of the powerful AGN population. The environment conditions, along with the larger number of powerful FRII sources at higher redshifts, would result in a relatively high efficiency of selection of HzRGs in surveys sampling high radio flux densities (above tens of mJy). On the other hand, at sub-mJy radio flux densities, the USS criteria will essentially only be able to select sources from the FRI and radio-quiet AGN population, as FRIIs are extremely rare (only $\sim3-4$ are predicted to be detected in the current work, irrespective of radio spectral index). The selection of pure star-forming galaxies, increasingly abundant at these faint radio flux densities, seems unlikely, given the synchrotron radio spectrum ($\alpha\sim-0.7$) they usually display. In fact, the redshift distribution of the USS sources in the current work (see Figure~\ref{fig:z-distr}, also reproduced in the appropriate panel of Figure~\ref{fig:models}) is remarkably similar to the  FRI and RQ redshift distribution from the $S^3$ simulations in Figure~\ref{fig:models} (Survey Model I), peaking just below $z\sim1$ and tailing-off above this redshift. The $\sim 40$ per cent USS sources in the Lockman Hole sample with no redshift estimates, if indeed at higher redshifts, would match well the model predictions, which place several tens of FRI sources at $z>2$ for this survey. Interestingly, the efficiency of the USS technique at sub-mJy flux densities for the selection of very high redshift galaxies may still be significantly high -- possibly even higher than that for radio surveys at much higher flux densities. Both Survey Models indicate similar numbers of detectable powerful AGN at $z\gtrsim 2$: around 80 FRIIs for Survey Model II and the same number of FRIs for Survey Model I. This raises the expectations for the application of the USS technique in the upcoming deepest LOFAR surveys, as it indicates the possibility of complementing the sampling of the high-redshift highest power radio sources (FRIIs) at high radio flux densities with the selection of co-eval lower power AGN population (FRIs) at sub-mJy levels. One should stress, however, that the uncertainties in the SKADS simulations, in particular for the population of FRIs at high redshifts, prevent us from drawing definite conclusions. Nevertheless, the importance of exploring the sub-mJy USS population with the faintest infrared magnitudes is obvious.

\section{Conclusions \label{sect:concl}}

We present a sample of 58 Ultra Steep Spectrum faint ($S_{\rm 610\,MHz}>100$\,$\mu$Jy) radio sources in the Lockman Hole. High reliability IRAC-SERVS 3.6\,$\mu$m counterparts are found for 48 (83\%) of these sources, with an average [3.6] total magnitude of 19.8\,mag (AB). Infrared colour-colour diagnostics indicate a significant fraction of IR-detected AGN among the radio-faint USS population. Using spectroscopic redshifts for 14 of the USS sources, complemented with photometric redshift estimates for a further 19 sources, the redshift distribution of USS sources is derived. Redshifts range from $z=0.1$ to $z=2.8$, peaking at $z\sim 0.6$ and tailing off at high redshifts. The remaining 25 sources include the faintest sources at IR wavelengths, and are potentially at higher redshifts. These results are essentially unchanged when the sample is restricted to the 30 sources (the USS(A) subsample) with the most secure USS classification, a procedure followed here to help identifying any bias from the small fraction (an estimated $\sim 25\%$) of non-USS sources scattered to a USS classification due to spectral index measurement errors. 

The comparison with the SKADS Simulated Skies models shows an indication that FRIs and RQ AGNs may constitute the bulk of the USS population at sub-mJy radio flux densities, and points to the possibility that the efficiency of this technique for the selection of higher redshift sources may be as high as when applied at much higher radio flux density levels. 

\acknowledgments  

JA, HM, MG, LB and CF gratefully acknowledge support from the Science and Technology Foundation (FCT, Portugal) through the research grant PTDC/FIS/100170/2008 and the Fellowships SFRH/BD/31338/2006 (HM) and SFRH/BPD/62966/2009 (LB).
JSD acknowledges the support of the Royal Society via a Wolfson Research Merit Award, and the support of the European Research Council via an ERC Advanced Grant.
CM and JP acknowledge support from the Marie Curie Excellence Team Grant
MEXT-CT-2006-042754 "UniMass" of the Training and Mobility of Researchers
programme financed by the European Community.

\clearpage
\pagestyle{plain} 
\begin{deluxetable}{c c rrr c cccc r ccc}
\tabletypesize{\scriptsize}
\rotate
\tablecolumns{13}
\tablewidth{0pc}
\tablecaption{Ultra Steep Spectrum Sources in the Lockman Hole \label{tab:sample}}
\tablehead{
\colhead{Source} & \colhead{flag} & \colhead{$S_{610}$} & \colhead{$S_{1400}$} & \colhead{$\alpha^{1400}_{610}$} & \colhead{$\alpha$} & \colhead{$\delta$} & \colhead{$\sigma_{pos}$} & \colhead{$\alpha_{SERVS}$} & \colhead{$\delta_{SERVS}$} & \colhead{[3.6]}   & \colhead{$\Delta^{radio}_{[3.6]}$} & \colhead{$z_{\rm spec}$}  &  \colhead{$z_{\rm phot}$} \\
\colhead{}       &                & \colhead{($\mu$Jy)} & \colhead{($\mu$Jy)}  & \colhead{}                      & \colhead{(J2000)}  & \colhead{(J2000)}  & \colhead{\arcsec}        & \colhead{(J2000)}          & \colhead{(J2000)}          & \colhead{(AB)}    & \colhead{\arcsec}                  & \colhead{}                & \colhead{}                \\
\colhead{(1)}       &     \colhead{(2)}        & \colhead{(3)} & \colhead{(4)}  & \colhead{(5)}                      & \colhead{(6)}  & \colhead{(7)}  & \colhead{(8)}        & \colhead{(9)}          & \colhead{(10)}          & \colhead{(11)}    & \colhead{(12)}                  & \colhead{(13)}                & \colhead{(14)}                \\
}
\startdata
 LH610MHzJ104949.1+571656  &   & 236  & 67      & $-1.51_{-0.44}^{+0.51}$ & 10:49:49.07 & +57:16:57.6 & 0.73 & 10:49:49.03 & +57:16:57.2 & $18.48$   & 0.47    & 0.351   & 0.400   \\[0.2ex]                 
 LH610MHzJ105004.3+572214  &   & 198  & 61      & $-1.40_{-0.36}^{+0.43}$ & 10:50:04.29 & +57:22:14.5 & 0.50 & 10:50:04.15 & +57:22:14.5 & $18.56$   & 1.12    & 0.244   & 0.400   \\[0.2ex]                 
 LH610MHzJ105004.4+572731  & A & 325  & 58      & $-1.58_{-0.34}^{+0.41}$ & 10:50:04.13 & +57:27:31.4 & 0.42 & 10:50:04.13 & +57:27:31.1 & $18.98$   & 0.30    & \nodata & \nodata \\[0.2ex]                           
 LH610MHzJ105009.2+571653  &   & 141  & 48      & $-1.30_{-0.46}^{+0.54}$ & 10:50:09.18 & +57:16:54.5 & 0.64 & 10:50:09.16 & +57:16:54.5 & $19.66$   & 0.19    & \nodata & \nodata \\[0.2ex]                           
 LH610MHzJ105025.4+572212  & A & 169  & \nodata & $< -1.63$               & 10:50:25.35 & +57:22:12.6 & 0.70 & 10:50:25.25 & +57:22:13.0 & $19.40$   & 0.88    & \nodata & 0.895   \\[0.2ex]                      
 LH610MHzJ105031.9+572027  & A & 5184 & 723     & $-2.37_{-0.08}^{+0.09}$ & 10:50:31.96 & +57:20:27.1 & 0.35 & 10:50:32.03 & +57:20:26.1 & $18.19$   & 1.14    & \nodata & 0.555   \\[0.2ex]                      
 LH610MHzJ105041.8+572130  &   & 144  & 44      & $-1.42_{-0.46}^{+0.47}$ & 10:50:41.66 & +57:21:31.1 & 0.51 &  \nodata    &  \nodata    & $> 22.71$ & \nodata & \nodata & \nodata \\[0.2ex]                                                     
 LH610MHzJ105042.2+572454  &   & 112  & 35      & $-1.38_{-0.45}^{+0.55}$ & 10:50:42.20 & +57:24:55.1 & 0.65 & 10:50:42.28 & +57:24:55.9 & $19.39$   & 1.05    & \nodata & 0.590   \\[0.2ex]                      
 LH610MHzJ105053.3+571549  &   & 137  &  \nodata & $< -1.36$ & 10:50:53.31 & +57:15:49.3 & 1.02 &  \nodata    &  \nodata    & $> 24.04$ & \nodata & \nodata & \nodata \\[0.2ex]                           
 LH610MHzJ105054.3+570416  & A & 243  & 68      & $-1.53_{-0.28}^{+0.31}$ & 10:50:54.28 & +57:04:16.3 & 0.31 & 10:50:54.25 & +57:04:16.9 & $19.30$   & 0.62    & \nodata & \nodata \\[0.2ex]                           
 LH610MHzJ105055.6+571803  &   & 142  & 47      & $-1.31_{-0.34}^{+0.34}$ & 10:50:55.55 & +57:18:04.7 & 0.36 & 10:50:55.50 & +57:18:05.0 & $22.10$   & 0.49    & \nodata & \nodata \\[0.2ex]                           
 LH610MHzJ105056.6+573219  &   & 189  & 56      & $-1.46_{-0.45}^{+0.48}$ & 10:50:56.55 & +57:32:19.3 & 0.50 & 10:50:56.52 & +57:32:19.3 & $20.13$   & 0.26    & \nodata & 1.000   \\[0.2ex]                      
 LH610MHzJ105057.5+572955  & A & 326  & 99      & $-1.43_{-0.19}^{+0.19}$ & 10:50:57.43 & +57:29:55.8 & 0.20 & 10:50:57.43 & +57:29:55.9 & $20.89$   & 0.10    & \nodata & 1.005   \\[0.2ex]                      
 LH610MHzJ105101.9+565750  & A & 688  & 121     & $-2.09_{-0.30}^{+0.37}$ & 10:51:01.80 & +56:57:50.0 & 0.42 &  \nodata    &  \nodata    & $> 23.27$ & \nodata & \nodata & \nodata \\[0.2ex]                                                     
 LH610MHzJ105105.0+571922  & A & 176  & 55      & $-1.40_{-0.30}^{+0.32}$ & 10:51:04.87 & +57:19:23.2 & 0.32 & 10:51:04.89 & +57:19:23.0 & $19.88$   & 0.22    & \nodata & 1.795   \\[0.2ex]                      
 LH610MHzJ105108.4+573344  & A & 374  & 97      & $-1.62_{-0.25}^{+0.27}$ & 10:51:08.25 & +57:33:44.9 & 0.34 & 10:51:08.29 & +57:33:45.2 & $19.36$   & 0.45    & 1.540   & 1.645   \\[0.2ex]                 
 LH610MHzJ105108.9+572534  & A & 136  & \nodata & $< -1.46$               & 10:51:08.91 & +57:25:35.0 & 1.45 & 10:51:08.80 & +57:25:35.5 & $21.45$   & 1.01    & \nodata & \nodata \\[0.2ex]                           
 LH610MHzJ105115.3+571820  & A & 213  & 55      & $-1.63_{-0.26}^{+0.26}$ & 10:51:15.19 & +57:18:20.6 & 0.25 & 10:51:15.20 & +57:18:20.8 & $19.11$   & 0.22    & \nodata & 0.505   \\[0.2ex]                      
 LH610MHzJ105133.3+571833  &   & 101  & \nodata & $< -1.32$               & 10:51:33.35 & +57:18:33.4 & 0.85 & 10:51:33.41 & +57:18:33.7 & $21.91$   & 0.56    & \nodata & \nodata \\[0.2ex]                           
 LH610MHzJ105134.5+573218  & A & 181  & 54      & $-1.46_{-0.37}^{+0.36}$ & 10:51:34.37 & +57:32:17.2 & 0.36 & 10:51:34.39 & +57:32:17.4 & $19.67$   & 0.29    & \nodata & \nodata \\[0.2ex]                           
 LH610MHzJ105146.8+572032  &   & 155  & 45      & $-1.48_{-0.42}^{+0.42}$ & 10:51:46.58 & +57:20:32.4 & 0.66 & 10:51:46.55 & +57:20:33.0 & $20.87$   & 0.63    & 0.981   & \nodata \\[0.2ex]                      
 LH610MHzJ105152.8+571347  &   & 226  & 76      & $-1.30_{-0.33}^{+0.34}$ & 10:51:52.73 & +57:13:47.1 & 0.63 &  \nodata    &  \nodata    & $> 21.79$ & \nodata & \nodata & \nodata \\[0.2ex]                                                     
 LH610MHzJ105155.2+570409  &   & 195  & 54      & $-1.55_{-0.43}^{+0.52}$ & 10:51:55.01 & +57:04:09.6 & 0.64 & 10:51:55.09 & +57:04:09.7 & $19.33$   & 0.65    & 0.421   & 0.405   \\[0.2ex]                 
 LH610MHzJ105158.0+571940  &   & 159  & \nodata & $< -1.30$               & 10:51:58.03 & +57:19:40.4 & 1.28 & 10:51:57.77 & +57:19:39.9 & $20.32$   & 2.15    & 1.368   & 2.080   \\[0.2ex]                 
 LH610MHzJ105200.3+571052  & A & 135  & \nodata & $< -1.43$               & 10:52:00.32 & +57:10:52.7 & 0.85 &  \nodata    &  \nodata    & $> 19.80$ & \nodata & \nodata & \nodata \\[0.2ex]                                                          
 LH610MHzJ105201.4+565947  & A & 2181 & 480     & $-1.82_{-0.26}^{+0.30}$ & 10:52:01.12 & +56:59:47.6 & 2.04 &  \nodata    &  \nodata    & $> 24.22$ & \nodata & \nodata & \nodata \\[0.2ex]                                                     
 LH610MHzJ105201.5+571105  & A & 177  & \nodata & $< -1.64$               & 10:52:01.49 & +57:11:06.0 & 0.95 & 10:52:01.29 & +57:11:08.1 & $18.90$   & 2.67    & \nodata & \nodata \\[0.2ex]                           
 LH610MHzJ105207.8+573346  &   & 170  & 57      & $-1.30_{-0.43}^{+0.39}$ & 10:52:07.66 & +57:33:46.8 & 0.36 & 10:52:07.67 & +57:33:46.6 & $19.18$   & 0.25    & 0.501   & 0.510   \\[0.2ex]                 
 LH610MHzJ105208.0+571349  &   & 153  & 48      & $-1.39_{-0.32}^{+0.31}$ & 10:52:07.91 & +57:13:49.6 & 0.28 &  \nodata    &  \nodata    & $> 22.62$ & \nodata & \nodata & \nodata \\[0.2ex]                                                     
 LH610MHzJ105208.1+571321  &   & 140  & 46      & $-1.33_{-0.43}^{+0.43}$ & 10:52:08.08 & +57:13:21.8 & 0.65 & 10:52:08.17 & +57:13:22.7 & $21.56$   & 1.16    & \nodata & \nodata \\[0.2ex]                           
 LH610MHzJ105208.3+571342  & A & 178  & \nodata & $< -1.48$               & 10:52:08.28 & +57:13:42.2 & 0.87 &  \nodata    &  \nodata    & $> 24.67$ & \nodata & \nodata & \nodata \\[0.2ex]                                                          
 LH610MHzJ105212.8+570641  & A & 215  & 56      & $-1.61_{-0.40}^{+0.48}$ & 10:52:12.63 & +57:06:41.3 & 0.72 & 10:52:12.71 & +57:06:41.7 & $19.09$   & 0.77    & \nodata & 0.695   \\[0.2ex]                      
 LH610MHzJ105214.0+571841  &   & 131  & 44      & $-1.32_{-0.42}^{+0.44}$ & 10:52:14.04 & +57:18:42.0 & 0.66 & 10:52:14.02 & +57:18:42.3 & $20.20$   & 0.39    & \nodata & \nodata \\[0.2ex]                           
 LH610MHzJ105220.6+573930  & A & 296  & 54      & $-2.03_{-0.30}^{+0.31}$ & 10:52:20.42 & +57:39:31.5 & 0.28 & 10:52:20.43 & +57:39:31.6 & $20.21$   & 0.16    & \nodata & 1.460   \\[0.2ex]                      
 LH610MHzJ105220.7+573408  &   & 208  & 61      & $-1.47_{-0.43}^{+0.41}$ & 10:52:20.76 & +57:34:08.6 & 0.42 & 10:52:20.73 & +57:34:08.7 & $18.63$   & 0.29    & \nodata & 1.100   \\[0.2ex]                      
 LH610MHzJ105225.6+571337  & A & 173  & 49      & $-1.52_{-0.34}^{+0.34}$ & 10:52:25.53 & +57:13:38.5 & 0.31 & 10:52:25.57 & +57:13:38.4 & $19.45$   & 0.37    & 0.468   & 0.660   \\[0.2ex]                 
 LH610MHzJ105229.1+574615  & A & 357  & 89      & $-1.67_{-0.44}^{+0.54}$ & 10:52:29.04 & +57:46:16.6 & 0.77 & 10:52:29.03 & +57:46:16.8 & $17.97$   & 0.23    & \nodata & 0.405   \\[0.2ex]                      
 LH610MHzJ105229.4+574009  &   & 158  & 45      & $-1.49_{-0.52}^{+0.55}$ & 10:52:29.17 & +57:40:09.4 & 0.64 & 10:52:29.16 & +57:40:09.6 & $22.75$   & 0.25    & \nodata & \nodata \\[0.2ex]                           
 LH610MHzJ105230.2+570024  & A & 486  & 110     & $-1.78_{-0.39}^{+0.54}$ & 10:52:30.16 & +57:00:24.3 & 0.80 & 10:52:30.06 & +57:00:24.4 & $19.17$   & 0.82    & \nodata & \nodata \\[0.2ex]                           
 LH610MHzJ105245.5+573745  & A & 291  & 84      & $-1.49_{-0.25}^{+0.24}$ & 10:52:45.39 & +57:37:45.9 & 0.21 & 10:52:45.39 & +57:37:45.9 & $19.34$   & 0.05    & 1.677   & 1.495   \\[0.2ex]                 
 LH610MHzJ105256.1+574148  &   & 257  & 86      & $-1.32_{-0.30}^{+0.31}$ & 10:52:56.07 & +57:41:48.6 & 0.31 & 10:52:56.04 & +57:41:47.7 & $19.13$   & 0.97    & \nodata & 1.795   \\[0.2ex]                      
 LH610MHzJ105257.8+573058  &   & 115  & 33      & $-1.49_{-0.60}^{+0.54}$ & 10:52:57.65 & +57:30:58.8 & 0.50 & 10:52:57.66 & +57:30:58.7 & $19.90$   & 0.13    & 2.100   & \nodata \\[0.2ex]                      
 LH610MHzJ105258.1+573616  & A & 246  & 71      & $-1.50_{-0.33}^{+0.32}$ & 10:52:58.11 & +57:36:16.9 & 0.39 & 10:52:58.19 & +57:36:17.2 & $20.16$   & 0.67    & \nodata & 1.780   \\[0.2ex]                      
 LH610MHzJ105301.5+573429  & A & 224  & 59      & $-1.60_{-0.33}^{+0.31}$ & 10:53:01.34 & +57:34:30.3 & 0.28 & 10:53:01.37 & +57:34:30.2 & $19.37$   & 0.28    & \nodata & 0.645   \\[0.2ex]                      
 LH610MHzJ105303.7+571616  & A & 235  & \nodata & $< -1.68$               & 10:53:03.65 & +57:16:17.0 & 0.65 &  \nodata    &  \nodata    & $> 22.93$ & \nodata & \nodata & \nodata \\[0.2ex]                                                     
 LH610MHzJ105308.4+573557  &   & 132  & 40      & $-1.42_{-0.47}^{+0.48}$ & 10:53:08.43 & +57:35:58.4 & 0.50 & 10:53:08.34 & +57:35:58.3 & $20.67$   & 0.77    & \nodata & \nodata \\[0.2ex]                           
 LH610MHzJ105310.5+572320  & A & 131  & 29      & $-1.81_{-0.43}^{+0.46}$ & 10:53:10.51 & +57:23:21.9 & 0.50 & 10:53:10.51 & +57:23:21.4 & $19.17$   & 0.46    & \nodata & 0.695   \\[0.2ex]                      
 LH610MHzJ105319.2+572108  & A & 321  & 108     & $-1.31_{-0.17}^{+0.17}$ & 10:53:19.18 & +57:21:08.7 & 0.18 & 10:53:19.27 & +57:21:08.5 & $21.19$   & 0.72    & 2.760   & \nodata \\[0.2ex]                      
 LH610MHzJ105323.2+571638  & A & 126  & \nodata & $< -1.35$               & 10:53:23.16 & +57:16:38.4 & 0.74 &  \nodata    &  \nodata    & $> 22.94$ & \nodata & \nodata & \nodata \\[0.2ex]                                                     
 LH610MHzJ105332.0+574145  &   & 264  & 79      & $-1.44_{-0.40}^{+0.40}$ & 10:53:32.16 & +57:41:44.2 & 0.46 & 10:53:32.12 & +57:41:44.4 & $18.49$   & 0.33    & 0.113   & 0.200   \\[0.2ex]                 
 LH610MHzJ105345.3+572328  & A & 158  & 44      & $-1.53_{-0.42}^{+0.40}$ & 10:53:45.18 & +57:23:29.7 & 0.36 & 10:53:45.18 & +57:23:29.3 & $20.81$   & 0.39    & \nodata & \nodata \\[0.2ex]                           
 LH610MHzJ105354.5+572340  & A & 154  & 40      & $-1.62_{-0.44}^{+0.46}$ & 10:53:54.31 & +57:23:40.0 & 0.42 & 10:53:54.38 & +57:23:39.9 & $19.41$   & 0.59    & \nodata & 0.705   \\[0.2ex]                      
 LH610MHzJ105358.1+574001  &   & 293  & 95      & $-1.36_{-0.42}^{+0.39}$ & 10:53:58.08 & +57:40:02.5 & 0.48 & 10:53:58.02 & +57:40:03.4 & $18.92$   & 0.99    & \nodata & 1.055   \\[0.2ex]                      
 LH610MHzJ105401.2+572842  &   & 192  & 64      & $-1.31_{-0.36}^{+0.35}$ & 10:54:01.00 & +57:28:43.7 & 0.36 & 10:54:01.01 & +57:28:43.8 & $19.44$   & 0.14    & \nodata & 0.605   \\[0.2ex]                      
 LH610MHzJ105407.0+573308  & A & 285  & 77      & $-1.56_{-0.36}^{+0.34}$ & 10:54:06.87 & +57:33:08.7 & 0.37 & 10:54:06.87 & +57:33:09.2 & $19.80$   & 0.50    & 2.560   & 1.710   \\[0.2ex]                 
 LH610MHzJ105429.1+572939  &   & 130  & 43      & $-1.32_{-0.52}^{+0.51}$ & 10:54:29.08 & +57:29:41.5 & 0.42 & 10:54:29.11 & +57:29:41.2 & $19.57$   & 0.40    & \nodata & 1.010   \\[0.2ex]                      
 LH610MHzJ105500.9+573344  &   & 253  & 75      & $-1.46_{-0.51}^{+0.50}$ & 10:55:01.06 & +57:33:45.9 & 0.50 & 10:55:01.02 & +57:33:45.9 & $19.89$   & 0.36    & \nodata & \nodata \\[0.2ex]                           
 LH610MHzJ105502.3+573224  &   & 297  & 98      & $-1.33_{-0.39}^{+0.40}$ & 10:55:02.27 & +57:32:25.9 & 0.44 & 10:55:02.29 & +57:32:26.0 & $18.80$   & 0.17    & 0.421   & 0.450   \\                 
\enddata
\tablecomments{The columns display (1) source name; (2) flag for sources included in a more robust sub-sample ``A"; (3) and (4) the flux densities at 610\,MHz and 1.4\,GHz, respectively; (5) radio spectral index and associated error; (6) and (7) radio position from the higher resolution 1.4\,GHz data, where available, and 610\,MHz otherwise; (8) error in the radio position; (9) and (10) the IR position of the SERVS-3.6\,$\mu$m counterpart; (11) 3.6\,$\mu$m counterpart total magnitude; (12) separation between the radio position and the IR counterpart; (13) and (14) the redshift measurement from spectroscopy or photometry measurements, respectively.}
\end{deluxetable}
\clearpage

\pagestyle{plaintop} 
\begin{deluxetable}{rcl}
\tabletypesize{\small}
\tablecolumns{3}
\tablewidth{0pc}
\tablecaption{SERVS identification rates of faint radio sources in the Lockman Hole \label{tab:ids}}
\tablehead{
\colhead{$\alpha^{1400}_{610}$} & \colhead{Number of radio sources} & \colhead{ID rate}
}
\startdata
$\alpha > -0.1$           & 40  & 98\% \\
$ -0.5<\alpha\leq -0.1 $  & 117 & 97\% \\
$ -0.9<\alpha\leq -0.5 $  & 273 & 92\% \\
$ -1.3<\alpha\leq -0.9 $  & 177 & 91\% \\
$\alpha\leq -1.3$         & 48  & 90\% \\
\enddata
\tablecomments{This table only considers sources with a spectral index determination, i.e., with a 1.4\,GHz detection. Those sources, and in particular USS sources, with only a 610\,MHz detection will have a lower radio positional accuracy, and, consequently, the probability of chance coincidences will be higher. Thus the overall identification rate for the general USS population in this sample, including ten 1.4\,GHz non-detections, is slightly lower.}.
\end{deluxetable}

\begin{deluxetable}{clllr}
\tablecolumns{5}
\tablewidth{0pc}
\tablecaption{Spectroscopic redshifts for USS sources in the Lockman Hole \label{tab:spectz}}
\tablehead{
\colhead{Name} & \colhead{$R$} & \colhead{[3.6]} & \colhead{$z_{spec}$} & \colhead{Reference for $z_{spec}$}\\
\colhead{} & \colhead{(AB)} & \colhead{(AB)} & \colhead{} & \colhead{}
}
\startdata
LH610MHzJ104949.1+571656  & 19.61   & 18.48 & 0.351 & \citet{Huang10}         \\
LH610MHzJ105004.3+572214  & 19.80   & 18.59 & 0.244 & \citet{Huang10}         \\
LH610MHzJ105108.4+573344  & 23.11   & 19.36 & 1.540 & \citet{Schmidt98}       \\
LH610MHzJ105146.8+572032  & \nodata & 20.87 & 0.981 & S. Chapman, priv. comm. \\      
LH610MHzJ105155.2+570409  & 19.96   & 19.34 & 0.421 & \citet{Huang10}         \\
LH610MHzJ105158.0+571940  & 23.61   & 20.32 & 1.368 & S. Chapman, priv. comm. \\           
LH610MHzJ105207.8+573346  & 21.71   & 19.18 & 0.501 & S. Chapman, priv. comm. \\ 
LH610MHzJ105225.6+571337  & 21.37   & 19.45 & 0.468 & \citet{Huang10}         \\
LH610MHzJ105245.5+573745  & 21.98   & 19.34 & 1.677 & \citet{Lehmann01}       \\
LH610MHzJ105257.8+573058  & \nodata & 19.90 & 2.100 & S. Chapman, priv. comm. \\      
LH610MHzJ105319.2+572108  & \nodata & 21.19 & 2.76  & \citet{Coppin10}        \\      
LH610MHzJ105332.0+574145  & 19.55   & 18.49 & 0.113 & \citet{Huang10}         \\
LH610MHzJ105407.0+573308  & 23.18   & 19.80 & 2.56  & \citet{Coppin10}        \\
LH610MHzJ105502.3+573224  & 20.43   & 18.80 & 0.421 & \citet{Huang10}         \\ 
\enddata
\tablecomments{The [3.6] and $R$-band magnitudes are total magnitudes.}
\end{deluxetable}

\clearpage

\pagebreak

\end{document}